\documentclass[preprint,12pt,3p,authoryear]{elsarticle}

\usepackage{aas_macros}
\usepackage[export]{adjustbox}
\usepackage{amssymb}
\usepackage{amsmath}
\usepackage{amsthm}
\usepackage{array}
\usepackage{colortbl}
\usepackage{color}
\usepackage{enumitem}
\usepackage{float}
\usepackage{footmisc}
\usepackage{graphicx}
\usepackage{hyperref}
\usepackage[utf8]{inputenc}
\usepackage{lineno}
\usepackage{longtable}
\usepackage{mathtools}
\usepackage{morefloats} 
\usepackage{multirow} 
\usepackage{rotating}
\usepackage{subcaption}
\usepackage{threeparttable}
\usepackage{url}
\usepackage{wasysym}
\usepackage[table]{xcolor}
\usepackage{wrapfig}

\newcommand{\fig}[2]{{fig.\,\ref{#1}{#2}}}
\newcommand{\tab}[1]{{table\,\ref{#1}}}
\newcommand{\eqn}[1]{{eq.\,\ref{#1}}}

\newcommand{\ith}{{$^{th}$}}
\newcommand{\eg}{{\it e.g.}}
\newcommand{\ie}{{\it i.e.}}

\newcommand{\dif}{\mathrm{d}}
\newcommand{\overbar}[1]{\mkern 1.5mu\overline{\mkern-1.5mu#1\mkern-1.5mu}\mkern 1.5mu}

\newcommand{\aei}{(a,e,i)}

\newcommand{\deltav}{\Delta v}

\newcommand{\NumPy}{{\texttt{NumPy}}}

\newcommand{\UnivariateSpline}{{\texttt{UnivariateSpline}}}

\newcommand{\arcdeg}{{^{\circ}}}

\newcommand{\dpd}{\,\mathrm{deg}\,\mathrm{day}^{-1}}

\newcommand{\au}{\,\mathrm{au}}
\newcommand{\km}{\,\mathrm{km}}

\newcommand{\kps}{\,\mathrm{km}\,\mathrm{s}^{-1}}

\newcommand{\meter}{\,\mathrm{m}}
\newcommand{\cm}{\,\mathrm{cm}}

\newcommand{\um}{\,\mu \mathrm{m}}

\newcommand{\yr}{\,\mathrm{yr}}
\newcommand{\Myr}{\,\mathrm{Myr}}

\newcommand{\Days}{\,\mathrm{days}}
\newcommand{\Day}{\,\mathrm{day}}
\newcommand{\days}{\,\mathrm{d}}

\newcommand{\rad}{\,\mathrm{rad}}

\newcommand{\second}{\,\mathrm{s}}
\newcommand{\mps}{\,\meter\,\second^{-1}}

\newcommand{\kg}{\,\mathrm{kg}}

\newcommand{\namedasteroid}[2]{{({#1})\,{#2}}}
\newcommand{\numberedasteroid}[3]{{#1}\,({#2}\,{#3})}

\newcommand{\VG}{{1991\,VG}}

\newcommand{\RH}{{2006\,RH$_{120}$}}
\newcommand{\CD}{{2020\,CD$_3$}}
\newcommand{\PT}{{2024\,PT$_5$}}
\newcommand{\NX}{{2022\,NX$_1$}}

\newcommand{\Kamo}{{\namedasteroid{469219}{Kamo‘oalewa}}}

\newcommand{\tIntegrationDays}{$2\times10^{10}\Days$}
\newcommand{\tIntegrationYears}{$54\Myr$}
\newcommand{\nMassiveBodies}{10}

\newcommand{\minimoonSaveRateDays}{0.1}
\newcommand{\heliocentricSaveRateDays}{7500}

\newcommand{\EarthMoonImpactRatio}{11.8}
\newcommand{\EarthMoonImpactRatioUnc}{1.8}
\newcommand{\DminMeters}{100}

\newcommand{\czero}{11.10}
\newcommand{\cone}{-8.61}
\newcommand{\ctwo}{3.03}
\newcommand{\cthree}{-0.44}
\newcommand{\VmodeAtDmin}{13.5}
\newcommand{\VmodeAtDonekm}{13.5}
\newcommand{\VmodeAtDmax}{12.2}
\newcommand{\MarchiDminMeters}{100}
\newcommand{\MarchiDmaxKms}{72}

\newcommand{\nCraters}{100}
\newcommand{\nSpeeds}{20}

\newcommand{\nEjectaPerSpeed}{6}

\newcommand{\EMSEscapeSpeedkps}{3.400}
\newcommand{\vEscapeClassicINmps}{2376}
\newcommand{\vEscapeHillINmps}{2343}
\newcommand{\minVkps}{3}
\newcommand{\maxVkps}{43}

\newcommand{\fLunarSImpactors}{   0.747}

\newcommand{\SdensitykgPerCubicMeter}{2700}
\newcommand{\SdensitykgPerCubicMeterUnc}{690}
\newcommand{\CdensitykgPerCubicMeter}{1410}
\newcommand{\CdensitykgPerCubicMeterUnc}{690}
\newcommand{\wgtdImpactorDensitykgPerCubicMeter}{2400}
\newcommand{\wgtdImpactorDensitykgPerCubicMeterUnc}{540}
\newcommand{\MoonDensitykgPerCubicMeter}{2550}
\newcommand{\craterDepthToDiameter}{0.2}
\newcommand{\EjectaDiameterMinMicrometers}{1}
\newcommand{\ejectaSFDslope}{3.7}
\newcommand{\ejectaSFDslopeRMS}{0.9}
\newcommand{\MaxPromptStartDay}{10}
\newcommand{\maxTimeBetweenSuccessiveCapturesDays}{10}

\newcommand{\gVpTenINkps}{9.6}
\newcommand{\gVpModeINkps}{13.5}
\newcommand{\gVpMedianINkps}{18.7}
\newcommand{\gVpNinetyINkps}{31.4}

\newcommand{\vZeroINkps}{2.380}

\newcommand{\nMinimoonsOneMeterIncremental}{30}
\newcommand{\nMinimoonsOneMeterCumulative}{36}

\newcommand{\collisionMinTimeDays}{1.1}
\newcommand{\fCollideParzero}{-0.015 \pm 0.003}
\newcommand{\fCollideParone}{0.06 \pm 0.02}
\newcommand{\fCollidePartwo}{-0.70 \pm 0.03}
\newcommand{\fCollideParthree}{-4.78 \pm 0.07}
\newcommand{\OneKmLunarImpactSpeed}{13.5}

\newcommand{\yearsGTCurrentOneMeterRate}{19000}

\newcommand{\VpModeOneHundredMeterImpactorINkps}{13.5}
\newcommand{\dCraterAtVModeOneHundredMeterImpactorINkm}{2.2}
\newcommand{\oneKmImpactorCraterDiameterKM}{13}
\newcommand{\oneKmImpactorCraterVolumeKMcubed}{176}
\newcommand{\oneKmImpactorEjectaGToneMeterInMillions}{4}
\newcommand{\maxDelayedCaptureFraction}{0.56}
\newcommand{\sCaptureFractionSpline}{0.01}

\newcommand{\maxAvgPromptCaptureEjectionSpeedMPS}{3038}
\newcommand{\promptQuadraticBreakSpeedMPS}{2765}
\newcommand{\promptDurationNinetyPercentileYears}{12.4}
\newcommand{\promptDurationNinetyNinePercentileYears}{65}
\newcommand{\delayedSpeedPlotOffsetINmps}{5}
\newcommand{\fLunarSImpactorsUnc}{0.032}
\newcommand{\nSystematics}{1147}
\newcommand{\nSystematicsGood}{263}
\newcommand{\sysNMin}{7.0e-05}
\newcommand{\sysNMax}{9.3e+07}
\newcommand{\sysNOrders}{12}
\newcommand{\sysNAvg}{51}
\newcommand{\CollinsSlopeAvg}{3.95}
\newcommand{\CollinsSlopeStd}{0.09}
\newcommand{\SingerSlopeAvg}{4.22}
\newcommand{\SingerSlopeStd}{0.09}
\newcommand{\HHSlopeAvg}{3.90}
\newcommand{\HHSlopeStd}{0.09}
\newcommand{\HNSlopeAvg}{4.22}
\newcommand{\HNSlopeStd}{0.09}
\newcommand{\BottkeSlopeAvg}{3.97}
\newcommand{\BottkeSlopeStd}{0.09}

\newcommand{\daysThousandXCurrentOneMeterRate}{310}

\newcommand{\GirodanoBrunoImpactorDiameterSinger}{380}
\newcommand{\GirodanoBrunoImpactorDiameterHH}{2300}

\newcommand{\LSSTexpTime}{30}
\newcommand{\LSSTTrailingStartDPD}{0.8}

\newcommand{\LSSTTrailingInverseDPD}{1.25}

\newcommand{\VlimLSST}{23.7}
\newcommand{\SurveyRateLimit}{10}

\newcommand{\TobservableLSST}{7}
\newcommand{\ResidenceTimeBinWidtha}{0.01}
\newcommand{\ResidenceTimeBinWidthe}{0.01}
\newcommand{\ResidenceTimeBinWidthi}{0.2}
\newcommand{\missingResidenceTimePercentSemiMajorAxis}{15}
\newcommand{\missingResidenceTimePercentInclination}{1}

\newcommand{\TBOMinimoonPercentDelayed}{18}

\newcommand{\TBOMinimoonRatioPercentLowSpeed}{83}
\newcommand{\TBOMinimoonRatioPercentHighSpeed}{35}

\journal{Icarus}

\newcommand{\gline}{\arrayrulecolor[gray]{0.8}\hline\arrayrulecolor{black}} 

\begin{document}

\begin{frontmatter}

\title{The steady state population of Earth's minimoons\\of lunar provenance}

\author[label1]{Robert Jedicke}
\ead{jedicke@hawaii.edu}
\author[label2]{Elisa Maria Alessi}
\author[label1,label3]{Naja Wiedner}
\author[label1]{\\Mehul Ghosal}
\author[label4]{Edward B. Bierhaus}
\author[label5,label6]{Mikael Granvik}






\address[label1]{Institute for Astronomy, University of Hawai`i, 2680 Woodlawn Drive, Honolulu, HI 96822, USA}
\address[label2]{Istituto di Matematica Applicata e Tecnologie Informatiche, Consiglio Nazionale delle Ricerche (IMATI-CNR), Via Alfonso Corti 12, 20133 Milano, Italy}
\address[label3]{Universität Tübingen, Geschwister-Scholl-Platz, 72074 Tübingen, Germany}
\address[label4]{Lockheed-Martin, Denver, CO}
\address[label5]{Department of Physics, P.O. Box 64, 00014 University of Helsinki, Finland}
\address[label6]{Asteroid Engineering Laboratory, Lule\aa{} University of Technology, Box 848, 981 28 Kiruna, Sweden}

\begin{abstract}
This work examines the plausibility of a lunar origin of natural objects that have a negative total energy with respect to the geocenter, \ie\ $E_T=$potential$+$kinetic energy$<0$, while they are within 3 Earth Hill radii ($R_H$), a population that we will refer to as `bound'.  They are a super-set of the informally named population of `minimoons' which require that the object make at least one orbit around Earth in a synodic frame rotating with Earth and that its geocentric distance be $<R_H$ at some point while $E_T<0$.  Bounded objects are also a dynamical subset of the population of Earth's co-orbital population, objects in a 1:1 mean motion resonance with Earth or, less specifically, on Earth-like orbits.  Only two minimoons have been discovered to date, \RH\ and \CD, while \PT\ and \NX\ meet our condition for 'bound'.  The likely source region of co-orbital objects is either the MB of asteroids, lunar ejecta, or a combination of both.  Earlier works found that dynamical evolution of asteroids from the MB could explain the observed minimoon population, but spectra of \CD\ and \PT\ and Earth co-orbital \Kamo\ are more consistent with lunar basalts than any MB asteroid spectra, suggesting that the ejection and subsequent evolution of material from the Moon's surface contributes to the minimoon and, more generally, Earth's co-orbital population. This work numerically calculates the steady-state size-frequency distribution of the bound population given our current understanding of the lunar impact rate, the energy of the impactors, crater-scaling relations, and the relationship between the ejecta mass and speed.  We numerically integrate the trajectory of lunar ejecta and calculate the statistics of `prompt' bounding that take place immediately after ejection, and `delayed' bounding that occurs after the objects have spent time on heliocentric orbits.  A sub-set of the delayed bound population composes the minimoon population.  We find that lunar ejecta can account for the observed population of bound objects but uncertainties in the crater formation and lunar ejecta properties induce a many orders of magnitude range on the predicted population.  If the bound objects can be distinguished as lunar or asteroidal in origin based on their spectra it may be possible to constrain crater formation processes and the dynamical and physical evolution of objects from the MB into near-Earth space.
\end{abstract}

\begin{keyword}
Asteroids \sep Asteroids, dynamics \sep Cratering \sep Moon \sep Near-Earth objects \sep Satellites, general
\end{keyword}

\end{frontmatter}



\newpage
\section{Introduction}
\label{s.Introduction}


`Minimoon' is a colloquial term for natural objects that complete an orbit around the Earth within a geocentric distance of 3 Earth Hill radii while being temporarily bound in the Earth-Moon system (EMS) with a negative geocentric Keplerian energy \citep{Kary1996-CaptureStatsOfSPCs,Granvik2012-minimoons}.  They must also be within one Earth Hill radii of the geocenter at some time during that period. The first discovered minimoon was the few-meter diameter \RH\ \citep{Kwiatkowski2008-2006RH120}.

\citet{Granvik2012-minimoons} and \citet{Fedorets2017-minimoons} modelled the minimoon population's size-frequency distribution (SFD) under the assumption that their provenance is the set of objects on Earth-like orbits, a subset of the near-Earth object (NEO) population, which mostly derive from the main belt (MB) between the orbits of Mars and Jupiter.  

This assumption has been challenged by spectroscopic studies of minimoons and objects closely related to them. \citet{Sharkey2021-2016HO3} reported that \Kamo, one of Earth's quasi-satellites, objects on heliocentric orbits in a 1:1 resonance with Earth, has an L-type spectrum resembling lunar silicates, and the minimoon \CD\ has a spectrum more consistent with the lunar surface than MB asteroids \citep{Bolin2020-CD3}.   \citet{Jedicke+2018Frontiers-Minimoons} were skeptical that lunar ejecta could be a minimoon source but \citet{Castro-Cisneros2023-LunarEjectaOriginOfKamo} dynamically modeled the evolution of material launched from the lunar surface after an impact and found that it is possible for lunar ejecta to evolve onto quasi-satellite orbits.  Similarly, this work models the production and dynamical evolution of lunar ejecta to calculate the SFD of minimoons and, more generally, temporarily bound objects (TBO) in the EMS. We define a TBO as any object with a negative geocentric Keplerian energy while within 3 Earth Hill radii of the geocenter.

The dynamical origin of Earth's minimoons with a MB provenance is similar to the temporary capture of comets and asteroids by the Jovian planets \citep[\eg][]{Carusi1981-TCOs}, with the most spectacular example being Comet Shoemaker-Levy 9's capture by Jupiter and subsequent impact into the giant planet \citep{Shoemaker1995-SL9,Kary1996-CaptureStatsOfSPCs}.  These temporarily captured objects are themselves a subset of the long-lived natural irregular satellite population of the Jovian planets. While the first known irregular satellite, Himalia, orbiting Jupiter, was discovered in 1904 \citep{Perrine1905-Himalia}, most of the other irregular satellites were discovered beginning in the late 20\ith\ century \citep[\eg][]{Gladman1998-UranusIrregularSatellites,Sheppard2003-JupiterIrregularSatellites}.

While \RH\ was Earth's first recognized minimoon discovery and \CD\ was the second \citep[\eg][]{Fedorets2020-2020CD3}, meteors have been detected that were likely minimoons before they entered Earth's atmosphere \citep[\eg][]{Clark2016-ImpactDetectionsTemporarilyCapturedNaturalSatellites}, including the spectacular meteor observed from Saskatchewan, Canada, to Bermuda in 1913 \citep{Chant1913a,Chant1913b}.  \citet{Granvik2012-minimoons} showed that about 1\% of minimoons become meteors while the others depart the EMS on Earth-like heliocentric orbits.  There have also been two identified TBOs, \PT\ \citep{FuenteMarcos2024-2024PT5} and \NX\ \citep{FuenteMarcos2023-minimoonsFromHorseshoes}.

Earth's minimoons and TBOs are of interest for many reasons.  They could be low $\deltav$ mission targets for returning asteroid samples to Earth and/or low-Earth orbit, with most minimoons having a sub-$\kps$ $\deltav$ from an Earth-Moon L$_2$ halo orbit \citep{Chyba2016-RH120,Jedicke+2018Frontiers-Minimoons}.  Their orbits could be modified to make their captures longer lived for more detailed study \citep[\eg][]{Urrutxua2015-TCO-retrieval}, or the orbit of TBOs could be modified to turn them into long-lived minimoons \citep[\eg][]{Baoyin2010,Brophy2012}.  While it is unlikely that a minimoon or TBO will be of a size and mineralogy that would make it commercially profitable, they represent a population of objects that could be utilized for testing in-situ resource utilization (ISRU) technology and operations in a low $\deltav$, rapid communication environment \citep[\eg][]{Granvik2013,Jedicke+2018Frontiers-Minimoons}.

TBOs also provide a scientific opportunity to test our understanding of the production and dynamical evolution of small objects from both the MB and the lunar surface.  Calculating the SFD of minimoons or TBOs with a main-belt provenance requires modeling the production of small asteroids as collision fragments in the MB, their Yarkovsky-driven migration into secular and mean-motion resonances, then their dynamical evolution into Earth-like orbits and subsequent capture in the EMS.  Similarly, minimoons or TBOs with a lunar-surface provenance require modeling the size and speed of the impactors, the crater formation process, the launch of material from the surface including its size-speed relationship, and then tracking the trajectories of the particles.  The particles must be propagated beyond the EMS because the objects can enter heliocentric orbit and be re-acquired as minimoons or TBOs years to millions of years in the future.  Thus, the TBO SFD and the ratio of lunar- to MB-generated TBOs could be a sensitive test of our understanding of all these processes.

Some recent studies have shown that lunar ejecta can evolve onto quasi-satellite orbits \citep{Castro-Cisneros2023-LunarEjectaOriginOfKamo,Jiao2024-Kamooalewa}.  In particular, \Kamo\ is an $\sim50\meter$ diameter asteroid with a spectrum consistent with being space-weathered lunar basalt.  Both papers suggest that not only did it originate from the Moon's surface but that it can be associated with the formation of the $\sim22\km$-diameter Giordano Bruno crater within the last 1 to $10\Myr$.  If this is the case then it is likely that many more quasi-satellites with lunar-like spectra will be identified in the future.  While quasi-satellite orbits are closely related to the orbits of objects that can become TBOs and minimoons, in the `classical' quasi-satellite regime analyzed by \citet{Castro-Cisneros2023-LunarEjectaOriginOfKamo} the particle orbits the Sun, but it appears to orbit Earth in retrograde motion in a reference system co-orbiting with Earth, without being captured by the Earth. The trajectories we consider are temporarily bound by the Earth and are at the same time co-orbital.

If lunar impacts can create quasi-satellites they should also produce smaller fragments that could become TBOs or minimoons. Recent observations of both \CD\ and \PT\ provide evidence to support that hypothesis because their spectra have the highest similarity to lunar surface material \citep{Bolin2024-PT5-arXiv,Kareta2024-2024PT5,Bolin2020-CD3}.  

The possibility that TBOs could be lunar ejecta was considered plausible by \citet{Tancredi1997-1991VG} based on dynamical arguments when examining the origin of \VG.  It had a EM barycentric eccentricity, $e<1$, identical to having a Keplerian energy $<0$, for about a month in early 1992 \footnote{JPL HORIZONS \url{https://ssd-api.jpl.nasa.gov/doc/horizons.htm}}, but did not meet the minimoon requirement of being within 3 $R_H$ at the time.  \VG\ was the first asteroid to be discovered on an orbit similar to Earth's and it generated speculation on whether it could be artificial or even an alien spacecraft \citep{Steel1995SETI-1991VG}.

\citet{Fedorets2020-2020CD3} considered the possibility that the minimoon \CD\ was launched from the Moon because some of their trajectory integrations included particles that intersected the lunar surface on 2017 September 15.  They concluded that this was an unlikely scenario because a lunar impact that could have launched a meter-scale asteroid on that date would have been noticed from Earth and detected by the Lunar Reconnaissance Orbiter (LRO) Narrow Angle Camera (NAC) \citep{Speyerer2016-NatureLROLunarCraterProduction}.

Thus, while there have been many studies of the dynamical evolution of lunar ejecta and their end states, relatively little attention has been placed on characterizing the geocentric portions of the ejecta's trajectories \citep[\eg][]{Gault1983-TerrestrialAccretionofLunarMaterial,Gladman+.1995.Icarus-LunarImpactEjecta,Castro-Cisneros2023-LunarEjectaOriginOfKamo}.  In this work we calculate the SFD of TBOs and minimoons of lunar provenance by modeling the different stages of their production and dynamical evolution.

\section{Method}
\label{s.Method}

This section details our algorithm for calculating the properties of the TBOs beginning with the defining equation for their steady state population (\S\ref{ss.Method_steady_state_SFD}), and then specifying our nominal choices for the lunar impactor size distribution (\S\ref{ss.FD}), the ejecta's size-speed distribution (\S\ref{ss.pDV}), the number density of lunar ejecta as a function of the impactor and ejecta size and speed (\S\ref{ss.n_ejecta}), the differential and cumulative lunar minimoon steady-state SFD (\S\ref{ss.SteadyStateSFD}) and, finally, a description of the dynamical integrations of the lunar ejecta (\S\ref{ss.LunarEjectaIntegrations}).  Throughout the description of our method we select input distributions and values from contemporary literature that reflect the field's current understanding, and implement specific distributions and values as our `nominal' scenario that yields the results provided in \S\ref{ss.minimoonSFD}.  The systematic uncertainties in our nominal scenario are then explored in \S\ref{ss.minimoonSFD_systematics}.

\subsection{The steady state lunar ejecta minimoon population}
\label{ss.Method_steady_state_SFD}

Our goal is to determine the steady-state number distribution of TBOs and minimoons that originated as lunar ejecta as a function of their diameter, $d$.  We will focus on the TBOs and later determine the fraction of TBOs that are also minimoons.  Let $\bar n(d) \, \dif d$ be the steady-state number of lunar ejecta TBOs in the diameter range $d \rightarrow d + \dif d$.  The number density of objects in the steady state is related to the average flux of objects, $\bar f(d)$, their net average rate of creation (or destruction since they are in the steady state), and their average lifetime, $\bar\ell(d)$:
\begin{equation}
    \bar n(d) = \bar f(d) \; \bar\ell(d).
    \label{eqn.nSteadyState-basic}
\end{equation}
We maintain the dependence of the flux and lifetime on an object's diameter because the rate of production of large objects in impacts is different from that of smaller objects. The lifetime of the objects could depend on diameter-dependent effects such as solar radiation pressure \cite[SRP, \eg][]{Vokrouhlicky2000-SRP} and Yarkovsky \citep[\eg][]{Vokrouhlicky2015-AsteroidsIV} but we will ignore these effects because they should be relatively small for the $d\ge1\meter$ objects and typically short TBO time frames under consideration.  Indeed, \eg, \citet{Langner2024-DART-Boulders}'s detailed dynamical simulation of meter-scale boulders in the Didymos-Dimorphos system confirmed that SRP is not an important consideration in this size and time domain. 

We also ignore the physical evolution of the objects during the integration, \eg\ comminution through collisions and tidal disruptions.  \citet{Bottke1996.Arjunas} showed that `[a]steroids which are collisionally decoupled from the main-belt', such as the objects considered here on Earth-like orbits, `catastrophically disrupt infrequently' and their Figure 7 suggests that the disruption probability is $\ll 10^{-9}\yr^{-1}$, \ie\ collisional lifetimes of $\gg 10^9\yr$.  This value agrees with the collisional lifetime of NEOs extrapolated from the MB collisional lifetimes in \citet{Bottke2005}.  That work suggests that MB asteroids in the $1\meter$ to $10\meter$ diameter range have collisional lifetimes of  $\sim10^7$ to $\sim3\times10^7$ years respectively.  A naive scaling of the MB collisional lifetime to NEOs based on their relative spatial densities suggests that the overall NEO collisional lifetime is $>10^9\yr$ and that of the `decoupled' objects considered here must be much longer.  Thus, we ignore collisional evolution because the dynamical lifetime of these objects of $\sim10-20\Myr$ \citep[\eg]{Bottke1996.Arjunas, Gladman+.1995.Icarus-LunarImpactEjecta} is about $100\times$ less than their collisional lifetime.

Tidal disruptions of NEOs may be responsible for an over-abundance of NEOs with semi-major axes close to $1\au$ \citep{Granvik2024-NEOTidalDisruption} but it is unlikely that tidal disruption of lunar-derived objects is important in this study. 
 \citet{Sridhar+Tremaine.1992.Icarus.Tidal_disruption_of_viscous_bodies}
 showed that a non-rotating, self-gravitating, viscous body will tidally
 disrupt if its periapse is $\lesssim 1.7 R_p
 (\rho_p/\rho_a)^{1/3}$, where $R_p$ and 
 $\rho_p$ are the planet's radius and density, respectively, and $\rho_a$ is the asteroid's density.  However, tidal disruption depends on many other factors as well, including the rotation period, shape and spin state of the asteroid, its bulk strength, and the speed at which the object is moving at periapse.  \citet{Richardson1998} studied the impact of these effects with numerical integrations and showed that slow encounter speeds with Earth, typical of objects on Earth-like orbits studied in this work, can disrupt asteroids in encounters at up to $\sim 3$ Earth-radii.  In our simulations of these types of encounters in \S\ref{ss.LunarEjectaIntegrations}, those that generate the lowest energy encounters with the EMS and produce minimoons, $\lesssim 2$\% of the capture events approach Earth within that distance, so we neglect tidal disruptions in this study.

Recognizing that the average flux and lifetime of the TBOs depends on the impact energy of the impactors that launch material from the lunar surface, the size-frequency and speed distributions of the ejecta, and the ejecta's probability of being bound, we expand on \eqn{eqn.nSteadyState-basic} and
\begin{enumerate}

    \item let the number of lunar impacts per unit time by objects with diameters, $D$, in the range $[D,D+\dif D]$ and speed, $V$, in the range $[V,V+\dif V]$ be $F'(D,V) \; \dif D \; \dif V = F(D) \; p(D,V) \; \dif D \; \dif V$ where $F(D)$ is the flux of lunar impactors of diameter $D$ and $p(D,V)$ is the probability density of an object of diameter $D$ impacting the lunar surface with speed $V$. \ie, we allow the impact speed distribution to depend on the impactor diameter. We employ capital letters to denote quantities related to the impactors and small letters to items concerning the ejecta/TBOs,
    
    \item let the number of ejected objects in the diameter range $[d,\dif d]$ and speed range 
    $[v,v+\dif v]$ that escape the Moon be $n_{escape}(D,V,d,v) \; \dif d \; \dif v$,
    
    \item let the average fraction of ejected objects with speed $v$ that ever become TBOs be $\bar f(v)$ and,
    
    \item let the average lifetime (duration) of lunar ejecta with speed $v$ as TBOs be $\bar\ell(v)$.
    
\end{enumerate}
Then
\begin{equation}
    \bar n(d) \; \Delta d = \; \Delta d \int \dif D \int \dif V \int \dif v
                       \; F(D) \; p(D,V) \; n_{escape}(D,V,d,v) \; \bar f(v) \; \bar\ell(v).
    \label{eqn.nSteadyState-full}
\end{equation}

The remainder of this section describes how we determine each of the five terms in the integrand of \eqn{eqn.nSteadyState-full}.  The terms are summarized in \tab{tab.eqn2termDefinitions} in \S\ref{ss.minimoonSFD_systematics}.

\subsection{The lunar impactor size distribution, $F(D)$}
\label{ss.FD}

\begin{figure}[htbp]
    \centering
    \includegraphics[width=0.48\columnwidth]{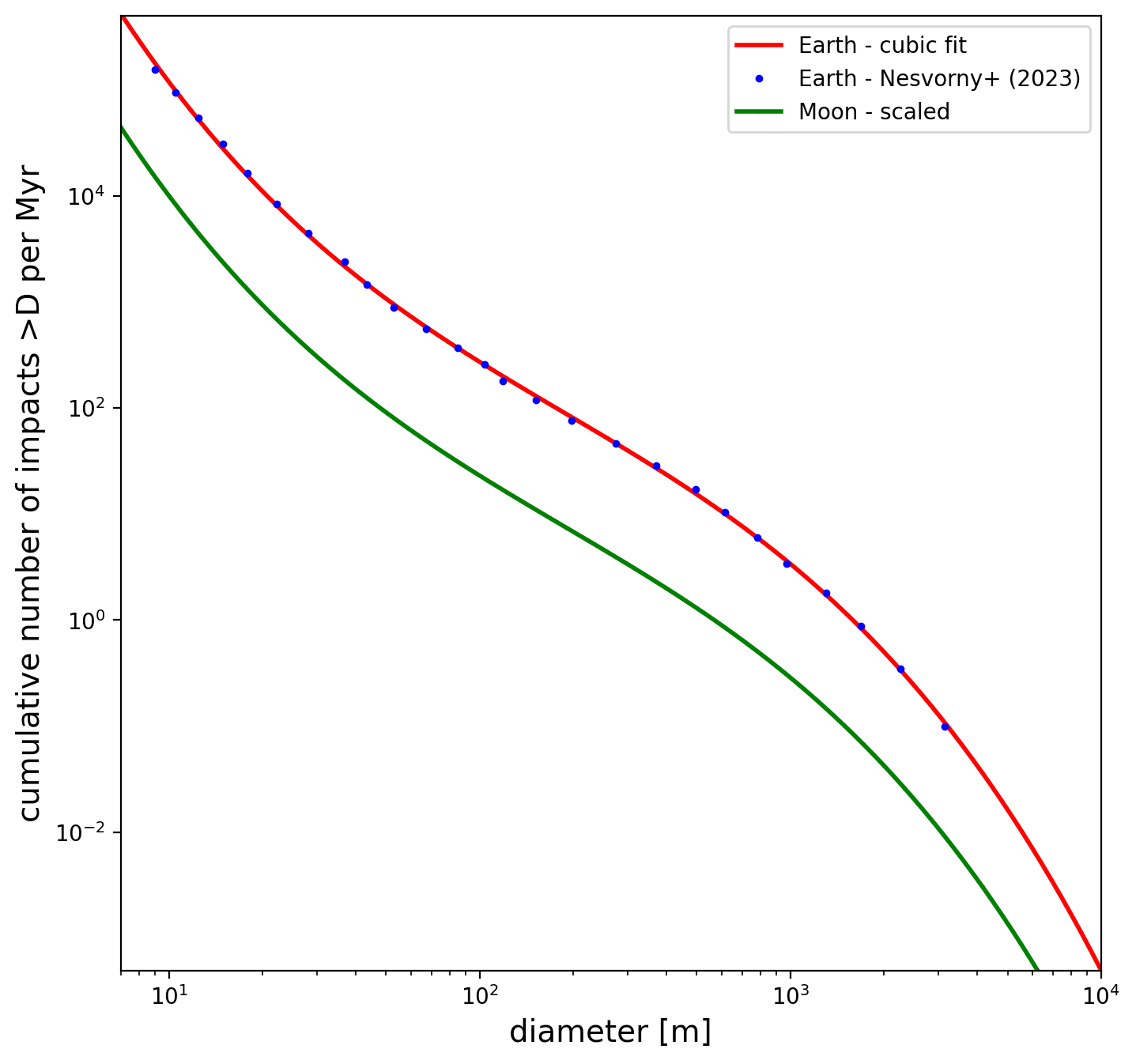}
    \includegraphics[width=0.48\columnwidth]{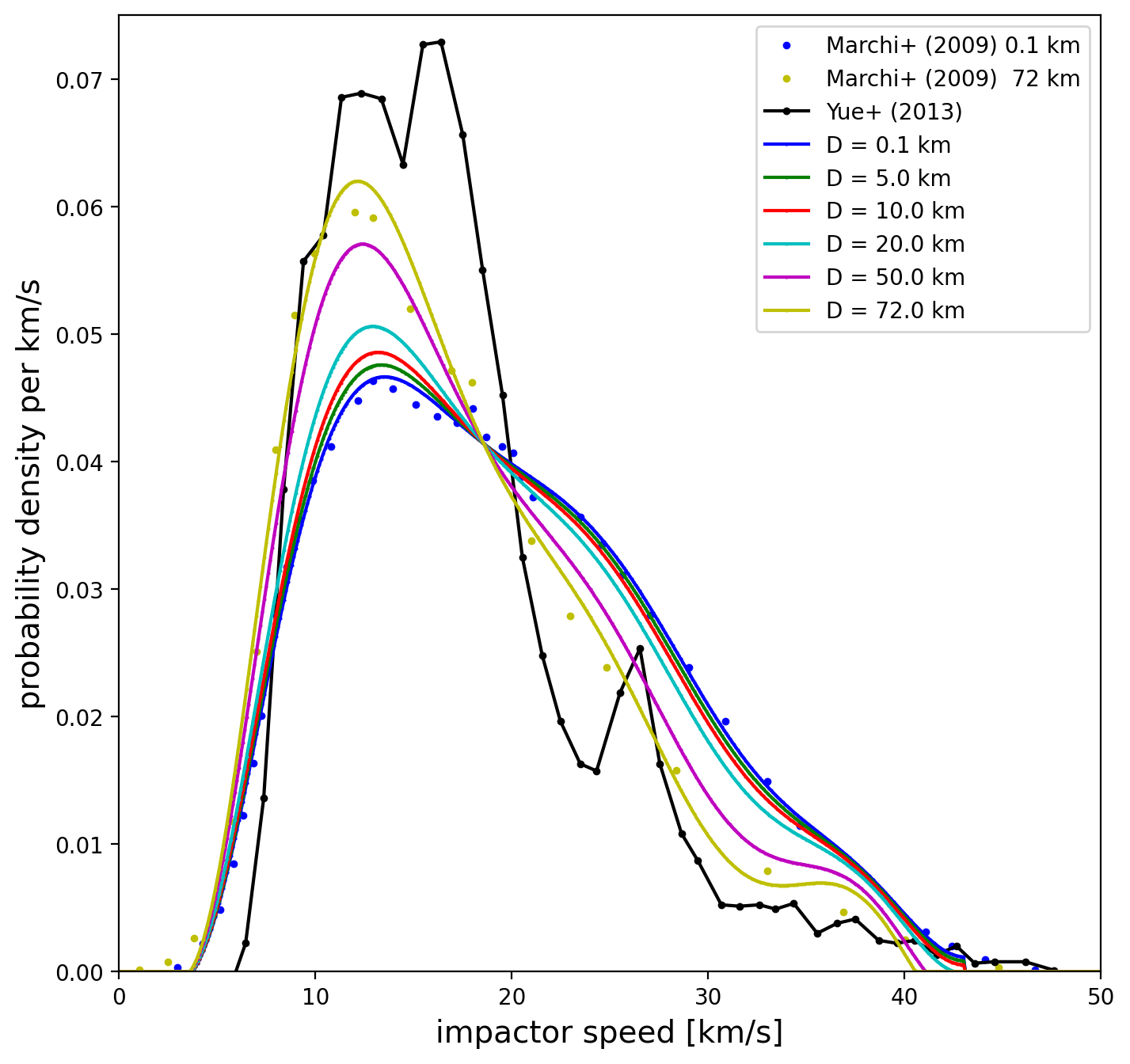}
    \caption{
        (left) The cumulative impact rate on Earth and the Moon vs. impactor diameter.  The blue points are digitized values from fig.~15 in \citet{Nesvorny2024-NEOMOD2}.
        (right) the lunar impact speed probability distribution for objects ranging from $\MarchiDminMeters\meter$ to $\MarchiDmaxKms\km$ diameter as adopted in this work from fits to digitized values from fig.~1 in \citet{Marchi-2009}.  The black points are digitized values from fig.~3 in \citet{Yue-2013}.}
    \label{fig.ImpactFlux+SpeedDistn}
\end{figure}

The impact rate on Earth for objects of $D\ge1\km$ diameter based on NEO statistics is $1.54\times10^{-6}\yr^{-1}$ \citep{Stokes2017,Nesvorny2024-NEOMOD2}.  That corresponds to an impact rate on Earth of about one $D\ge1\km$ diameter object every $650,000\yr$.  \citet{Werner2002-LunarImpactorSFD} state that the lunar impact rate of the same size objects is $(1.3\pm0.2)\times10^{-7}\yr^{-1}$ based on lunar crater counts and crater-scaling relations.  The ratio between the impact rates of $R_{E:M}=\EarthMoonImpactRatio\pm\EarthMoonImpactRatioUnc$ is in agreement with the ratio of the cross-sectional area of the Earth and Moon of about 13.4.  Thus, we used the Earth impactor SFD from \citet{Nesvorny2024-NEOMOD2} scaled down by a factor of \EarthMoonImpactRatio\ as our nominal lunar impactor SFD (\fig{fig.ImpactFlux+SpeedDistn}, left).

We fit the \citet{Nesvorny2024-NEOMOD2} cumulative Earth impact rate, $N_E(D)$, to a function of the form 
\begin{equation}
   \log N_E(D) = \sum_0^3 c_k \Bigg[ \log \bigg[ \frac{D}{\meter} \bigg] \Bigg]^k 
\end{equation}
where $c_k \sim (\czero,\cone,\ctwo,\cthree)$ for $k=0,3$.  The cumulative lunar impact rate is then $N_M(D) = N_E(D) / R_{E:M}$.  The lunar impactor flux, the differential impact rate on the Moon as a function of diameter only, is then $F(D) = \dif N_M(D) / \dif D$.

\subsection{The size-dependent lunar impactor speed distribution, $p(D,V)$}
\label{ss.pDV}

\citet{Marchi-2009} provide the speed distribution of lunar impactors at diameters of $D_1=100\meter$ and $D_2=72\km$ (\fig{fig.ImpactFlux+SpeedDistn}, right).  \citet{Yue-2013} provide a narrower lunar impact speed distribution with a maximum centered at about $14\kps$ and a secondary peak at $\sim 26.5\kps$, but they point out that their distribution `agree[s] closely' with \citet{Marchi-2009} and we chose to use the Marchi distributions because they enabled us to incorporate an impactor diameter dependence.  

We fit both \citet{Marchi-2009} distributions to a 10$^{th}$ order polynomial,
\begin{equation}
    p'(D_i,V) = \sum_{j=0}^{10} a'_{ij} \; V^j,
\end{equation}
truncated the distributions so that $p(D_i,V)=0$ for $V<\minVkps$ and $V>\maxVkps$ and $i=1,2$, and then normalized the distributions' coefficients ($a_{ij}$) so that $\int p(D_i,V) \; \dif V = 1$.  A 10th-order polynomial was the simplest model that empirically fit the shape of both data sets.  The functions were truncated because the fits behaved non-physically at the smallest and highest speeds and did not represent the data.  Finally, we calculate the impactor speed probability distribution at any diameter between $D_1$ and $D_2$ by interpolating the polynomial co-efficients such that the probability density of an impactor of diameter $D$ having an impact speed $V$ is
\begin{equation}
    p(D,V) = \sum_{j=0}^{10} \bigg[ a_{1j} + \frac{D-D_1}{D_2-D_1} ( a_{2j} - a_{1j} ) \bigg] \; V^j.
\end{equation}

The mode of the impact speed distribution (\fig{fig.ImpactFlux+SpeedDistn}) decreases slowly from $\sim\VmodeAtDmin\kps$ for impactors of $D=0.1\km$ diameter to $\sim\VmodeAtDmax\kps$ at $D=\MarchiDmaxKms\km$.

\subsection{The number density of lunar ejecta as a function of the impactor and ejecta size and speed, $n_{escape}(D,V,d,v)$}
\label{ss.n_ejecta}

When the lunar surface is struck by a large enough impactor some of the material ejected in the formation of the impact crater may be launched at a speed greater than the lunar escape speed but still be bound in the EMS, \ie, become a TBO, while faster material can escape the EMS into heliocentric orbit.  The classical formula for the escape speed from an isolated body yields a lunar escape speed of $\vEscapeClassicINmps\mps$, but \citet{Alvarellos2002,Alvarellos2005} clarified that the escape speed from the satellite of a planet is the speed required to reach the satellite's Hill sphere.  For the case of the Moon orbiting Earth, their formulation reduces the lunar escape speed by about 1.5\%, to $v_{esc}=\vEscapeHillINmps\mps$.

\begin{figure}[htbp]
    \centering
    \includegraphics[width=0.48\columnwidth]{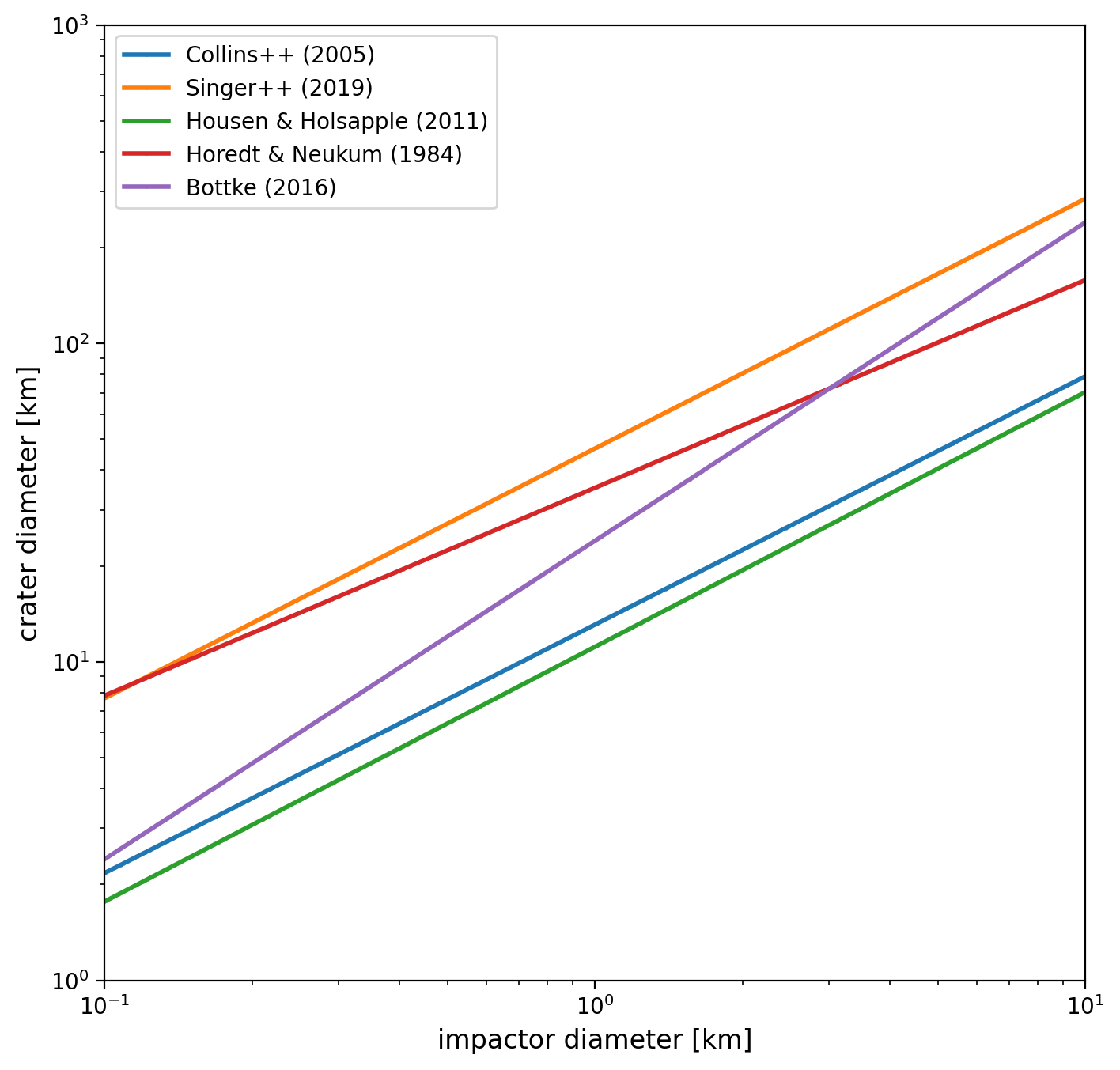}
    \includegraphics[width=0.48\columnwidth]{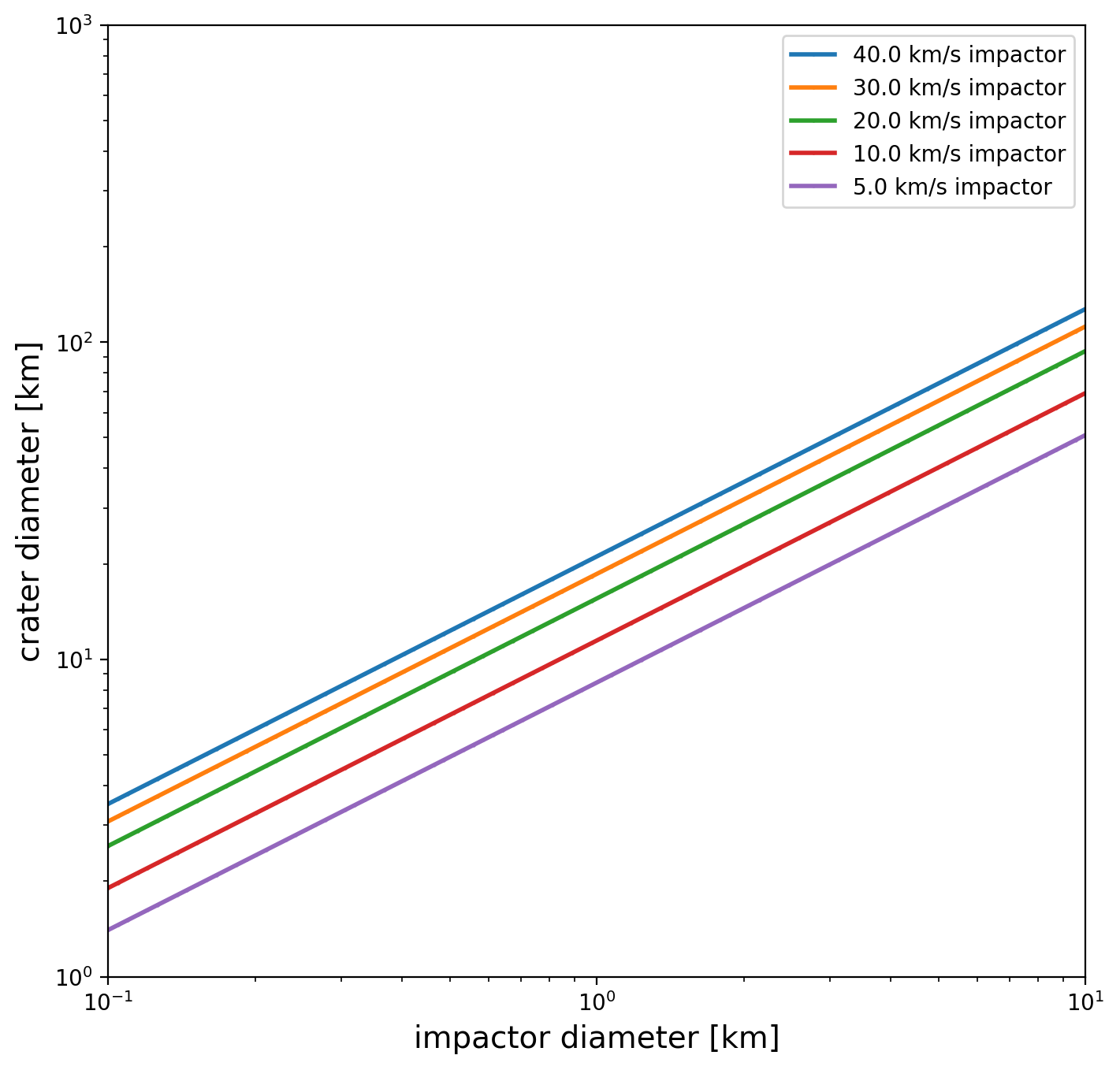}
    \caption{
        (left) Various crater scaling relationships proposed over the last forty years for lunar impacts with parameters as described in the text and an impactor speed of $13.5\kps$. We adopt the \citet{Collins2005-craterScaling} as our nominal parameterization for this work.
        (right) The relationship between crater diameter and the impactor's speed and diameter.}
    \label{fig.CraterScalingRelations}
\end{figure}

To determine the volume of material ejected during an impact we employ a crater-scaling relation that provides the size of an impact crater as a function of various parameters including the impactor's speed and mass.  Crater scaling relations are generally separated into two categories, the `strength regime' and `gravity regime', based on how the impact process and crater formation depend on the physical properties of the target material (\eg, the lunar surface) and the impactor (\eg, an asteroid).  The crater diameter at which the transition occurs between the regimes can be determined by the crater size at which their morphology switches from `simple' to `complex', at about $16\km$ diameter on the Moon \citep{Sun2023-LunarImpactCraters}.  The impactor diameter capable of creating a crater of this dimension ranges from about $200\meter$ to $2\km$ depending on which crater scaling relation is employed (\fig{fig.CraterScalingRelations}).   We will show below that the minimum diameter impactor that can \emph{just} liberate meter-scale TBOs from the Moon's surface is $\sim100\meter$ but the number of ejecta of $>1\meter$ diameter launched faster than the escape speed increases dramatically with the impactor diameter. Given the uncertainty on the impactor diameter that corresponds to the onset of the gravity regime, that the number of escaping ejecta is strongly dependent on impactor diameter, and to simplify our calculation, we assume that all the craters that launch TBOs are in the gravity regime.

Laboratory, numerical, and observational studies on crater scaling and ejecta size-speed relations are continually evolving \citep[\eg][]{Anderson2004-ejectionAngles,Hirase2004P-EjectaSizeVelocity,Bart2010-BouldersEjectedFromLunarCraters,Krishna2016-LunarImpactSpallation,Singer2020-EjectaBlockSizes} with modern crater scaling laws converging on agreement \citep[\fig{fig.CraterScalingRelations},][]{Johnson2016-CraterScaling}.  All the work is broadly consistent in the sense that 1) ejection fragment size decreases as ejection speed increases, 2) there is a correlation between  crater diameter and the maximum diameter of ejected fragments at a given ejection speed and, 3) increasing crater diameter results in increasing maximum ejection speed at a given fragment diameter, but \citet{Singer2020-EjectaBlockSizes} introduced a nuanced aspect showing that a fall-off in maximum ejection speed is a function of crater diameter. 

Given the broad agreement in impact scaling relations we adopted the one developed by \citet{Collins2005-craterScaling} for the diameter of a transient\footnote{We use the transient crater diameter rather than the final crater diameter because the impact launches material from the surface during the formation of the transient crater.} impact crater, 
\begin{equation}
    d_{crater}(D,V) 
      = 1.161 \; \Bigl (\frac{\rho_i}{\rho_t} \Bigr )^{1/3}  \;
        D^{0.78} \; V^{0.44} \; g^{-0.22} \; \sin^{1/3}\theta
    \label{eqn.CraterScaling}
\end{equation}
where $\rho_i$ is the density of the impactor, $\rho_t$ is the density of the target (lunar crust), $g$ is the gravitational acceleration at the target's surface. \ie, the Moon's, with $g=Gm/r^2$, where $G$ is the gravitational constant, $m$ is the impactor's mass, $r$ is its radius, and $\theta$ is the angle between the velocity vector of the impactor and the normal to the target's surface at the point of impact.  We use the most probable impact angle of $\theta=45\arcdeg$, as is common in cratering studies.

\citet{Carry2012} provide the average density\footnote{We used their $\rho_{50}$ values derived from objects with $<50$\% uncertainty on the density.  The values differ by $<11$\% from their $\rho_\infty$ and $\rho_{20}$ values.} for C- and S-class asteroids of $\rho_C = (\CdensitykgPerCubicMeter \pm \CdensitykgPerCubicMeterUnc) \kg\meter^{-3}$ and $\rho_S = (\SdensitykgPerCubicMeter \pm \SdensitykgPerCubicMeterUnc) \kg\meter^{-3}$, respectively.  We use the weighted average impactor density of $\rho_i = (\wgtdImpactorDensitykgPerCubicMeter \pm \wgtdImpactorDensitykgPerCubicMeterUnc) \kg\meter^{-3}$ assuming that the fraction of impactors that are S-class in the NEO population\footnote{The uncertainty on the ratio is much less than the uncertainty on the S- and C- class asteroid densities.} is $f_S=\fLunarSImpactors$ \citep{Wright2016}.  GRAIL observations provide an average lunar crust density of $\rho_t=\MoonDensitykgPerCubicMeter \kg\meter^{-3}$ \citep{Wieczorek2013-GRAIL-lunarCrust}.  

Given a crater diameter and shape we can determine the volume of material excavated in an impact and determine the SFD of the ejecta. 

\citet{Pike1974-LunarDepthDiameterRelations} used Apollo photogrammetry to show that the depth/diameter ratio for lunar craters of $<15\km$ diameter is $\Re\sim\craterDepthToDiameter$ ``but larger craters are not much deeper''.  Our crater-scaling relation suggests that a $15\km$ diameter lunar crater is formed by the impact of an $\sim1.2\km$ impactor at $13.5\kps$.  Since the largest known potentially hazardous object\footnote{Potentially hazardous objects have a minimum orbital intersection distance with Earth's orbit of $<0.05\au$ and an absolute magnitude $<22$.}, \numberedasteroid{53319}{1999}{JM8}, is $\sim7\km$ in diameter we employ $\Re$ as a constant.  We then assume that a lunar crater is a paraboloid of revolution so that its volume is
\begin{equation}
    V_c(D,V) =  \frac{\pi}{8} \, \Re \, \big[ d_{crater}(D,V) \big]^3\,,
    \label{eqn.Vcrater}
\end{equation}
\noindent and we explicitly show the crater's diameter dependence on the impactor's diameter and speed.

We then assume that the ejecta have a cumulative SFD of the form 
\begin{equation}
    N(>d) = C \, d^{-p_e}
    \label{eqn.Nd}
\end{equation}
and \citet{Bart2010-BouldersEjectedFromLunarCraters} provide values for $p_e$ for decameter scale boulders ejected from 18 lunar craters with an average $\overbar{p_e} = \ejectaSFDslope$ and a standard deviation of $\ejectaSFDslopeRMS$.  Adopting $\overbar{p_e}$ as the nominal value for our study, and recognizing that the ejecta volume is dominated by the smallest particles if $p_e>3$ \citep[\eg][]{Bierhaus2018}, we set $d_*=\EjectaDiameterMinMicrometers\um$ as the smallest reasonable ejecta size.
The total volume of material in the ejecta is then
\begin{eqnarray}
    V_e &=& \frac{\pi}{6} \, C \, p_e \, \int_{d_*}^{d_{max}} d^{-p_e-1} d^3 \dif d  \\
        &=& \frac{\pi}{6} \frac{C \, p_e}{3-p_e} \bigg| \; d_{max}^{3-p_e} - d_*^{3-p_e} \; \bigg|.
    \label{eqn.Vejecta}
\end{eqnarray}

The maximum ejecta diameter is given by $N = 1 = C \; d_{max}^{-p_e}$ so $d_{max} = C^{1/p_e}$. Equating the volume of the crater and ejecta, \eqn{eqn.Vcrater} and \eqn{eqn.Vejecta} respectively, and substituting $d_{max}$ yields an equation for the SFD normalization constant, $C$:
\begin{equation}
    C \big| C^{(\frac{3}{p_e}-1)} - d_*^{3-p_e} \big| =  \frac{3}{4} \, \bigg( \frac{3}{p_e}-1 \bigg) \, \Re \; [d_{crater}(D,V)]^3,
\end{equation}
where it is important to note that $C$ depends on the impactor's diameter and speed, \ie, $C \equiv C(D,V)$.

\begin{center}
    \begin{minipage}{10cm}  
        \includegraphics[width=\columnwidth]{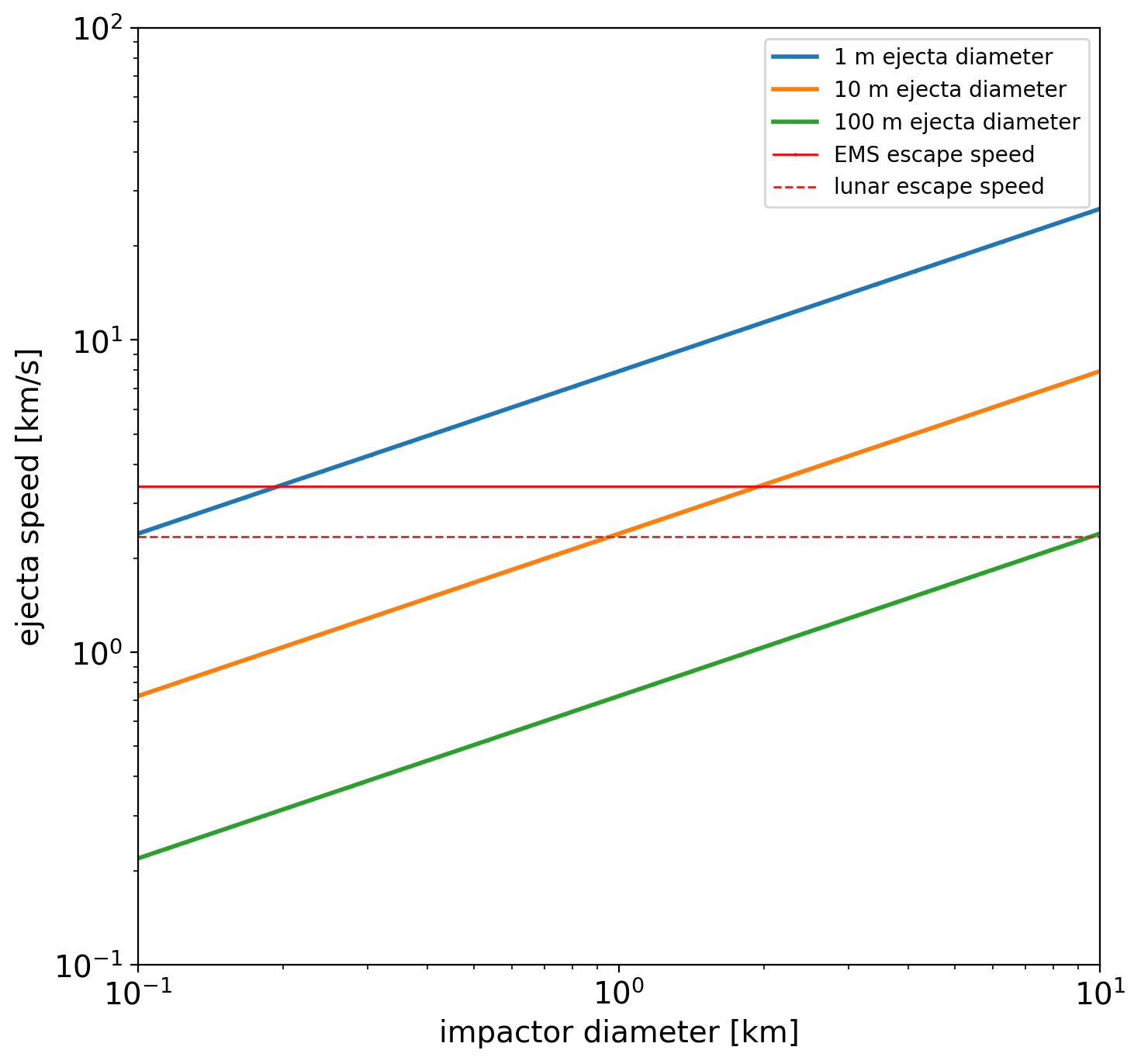}
        \captionof{figure}{
            Ejecta speed for 3 different ejecta diameters as a function of impactor diameter.  Ejecta can only escape the Moon's Hill sphere if they have speeds greater than the escape speed illustrated by the dashed red horizontal line.  Ejecta with speeds $>\EMSEscapeSpeedkps\kps$ as indicated by the solid red line must escape the EMS \citep{Gladman+.1995.Icarus-LunarImpactEjecta}.
        }
        \label{fig.EjectaSpeed}
    \end{minipage}
\end{center}

The final key to defining $n_{escape}(D,V,d,v)$ is the relationship between the ejecta diameter and speed.  We intuitively expect that the ejecta speed will be inversely related to the ejecta diameter, \ie\ bigger ejecta have slower speeds, as described below.  \citet{Hirase2004P-EjectaSizeVelocity} (fig.~4) provide data and formulae for lunar ejection speed vs. the ejecta/impactor size ratio ($R$) which we employ due to its simplicity despite the availability of more involved formulae \citep[\eg][]{Singer2020-EjectaBlockSizes}.  We approximate the \citet{Hirase2004P-EjectaSizeVelocity} (revised) best-case scenario for launching lunar material, \ie\ the version that provides the highest ejection speed at each $R$ so as to provide an optimistic number of objects that reach lunar escape speed, with the relationship:
\begin{equation}
    \log_{10} \bigg[ \frac{v}{\mps} \bigg] 
        = 2.34 - 0.52 \log_{10} \bigg[ \frac{d}{D} \bigg].
    \label{eqn.Hirase}
\end{equation}
Their results are provided in the range $0.05 \lesssim R \lesssim 0.7$ but we assume that the relationship can be extended to both smaller and larger $R$.  The minimum impactor diameter that can eject a $1\meter$-diameter fragment at the lunar escape velocity is about $D_{min}=\DminMeters\meter$ where $R=0.01$ (\fig{fig.EjectaSpeed}).  Our crater-scaling relationship (\eqn{eqn.CraterScaling}) suggests that such an impact would generate an $\sim\dCraterAtVModeOneHundredMeterImpactorINkm\km$-diameter crater at the most likely impact speed of $\VpModeOneHundredMeterImpactorINkps\kps$.  

\begin{figure}[htbp]
    \centering
    \includegraphics[width=0.48\columnwidth]{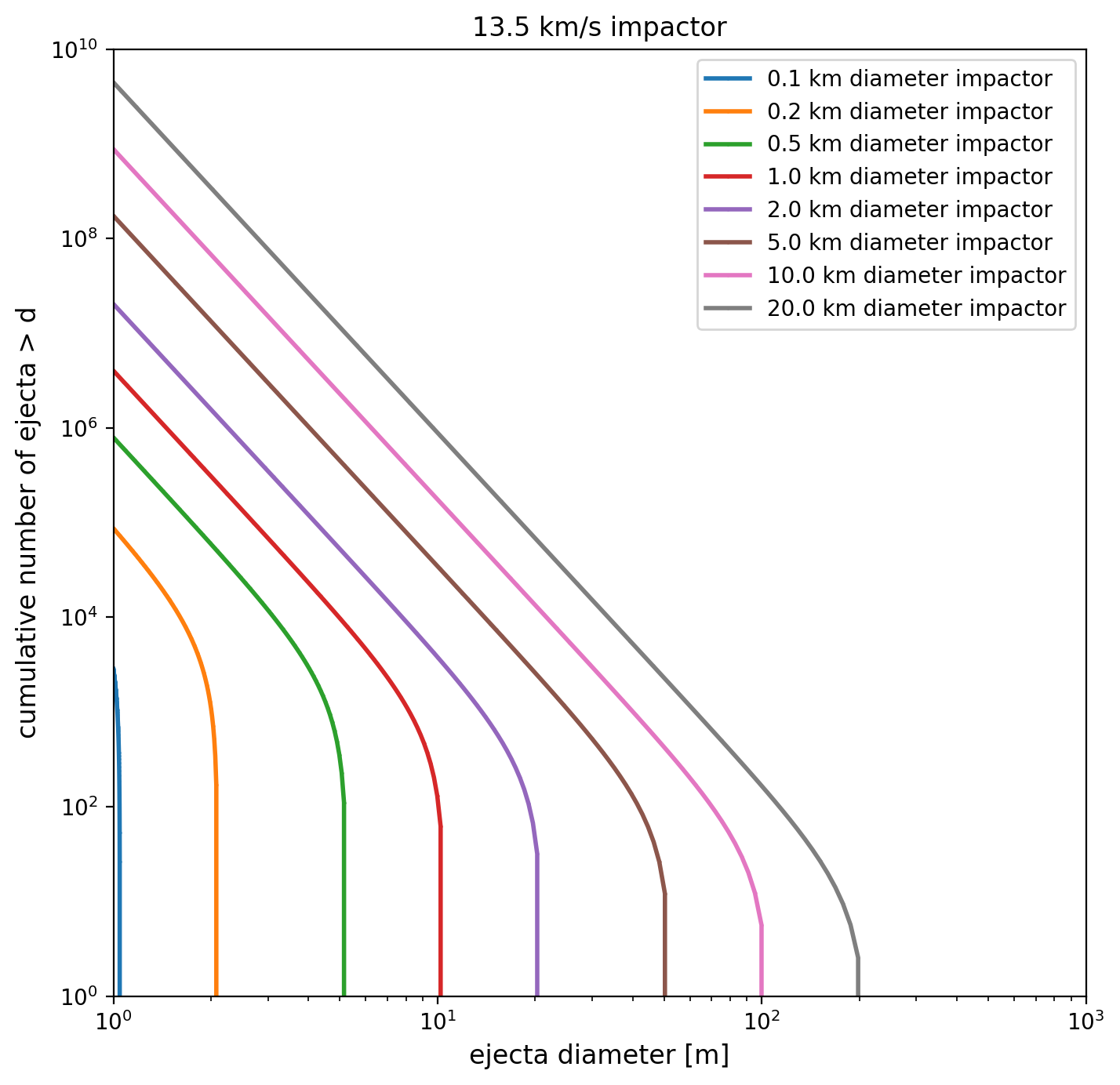}
    \includegraphics[width=0.48\columnwidth]{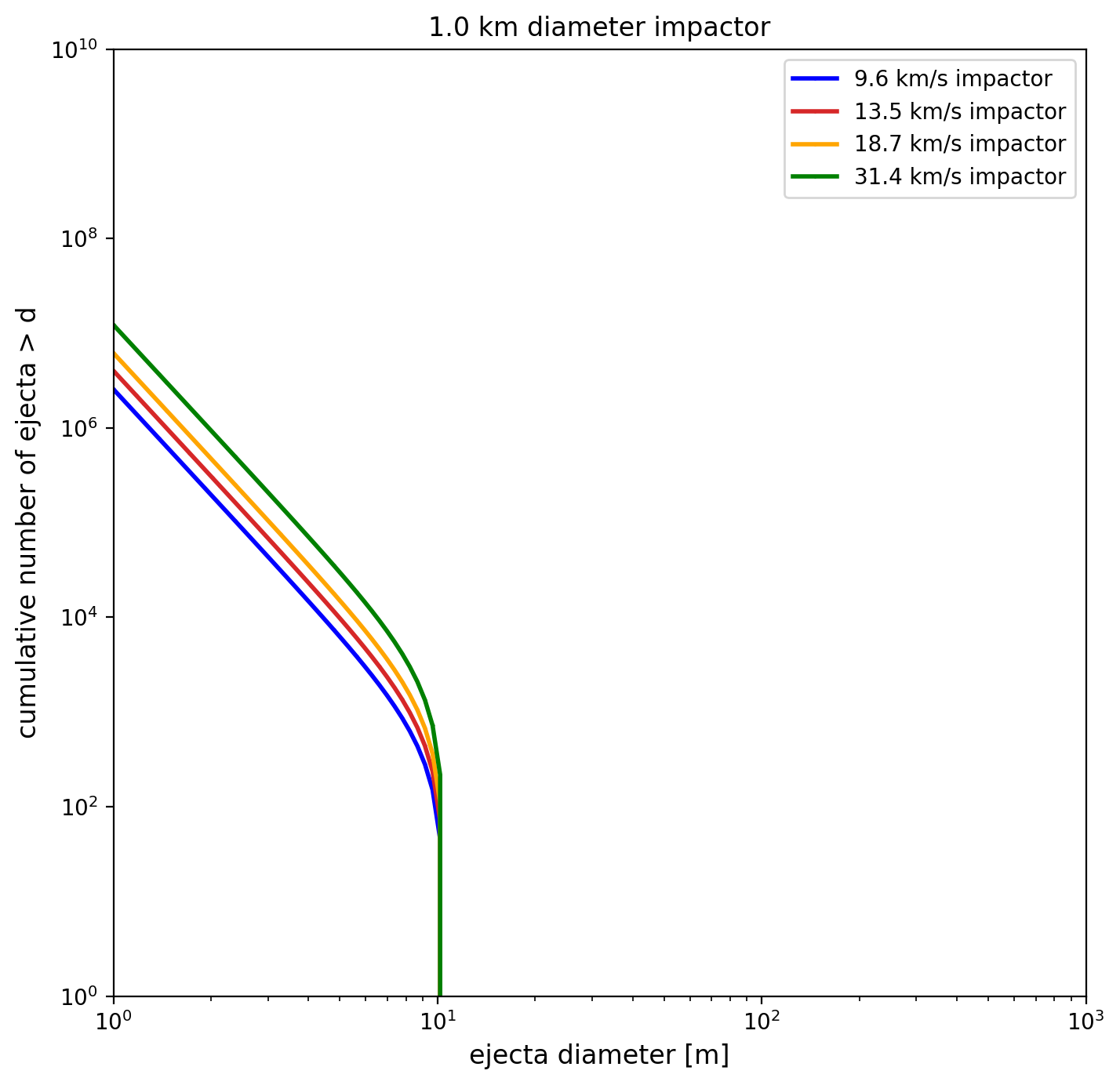}
    \caption{
        (left) The ejecta's cumulative SFD for objects larger than 1 meter in diameter for eight different impactor diameters colliding with the lunar surface at $\gVpModeINkps\kps$, close to the mode of the speed distribution. 
        (right)  The ejecta's cumulative SFD for objects larger than 1 meter in diameter for a $1\km$ diameter impactor at $\gVpTenINkps\kps$, $\gVpModeINkps\kps$, $\gVpMedianINkps\kps$ and $\gVpNinetyINkps\kps$, close to the 10\%, mode, median, and 90\% impact speeds.}
    \label{fig.ejectaSFD}
\end{figure}

Thus, we arrive at the ejecta's differential SFD, similar in form to \eqn{eqn.Nd}, as a function of the impactor's and ejecta's diameter and speed (\fig{fig.ejectaSFD}): 
\begin{eqnarray}
    n_{escape}(D,V,d,v) &=& 
    \begin{cases}
        C(D,V) \, p_e \, d^{-p_e-1}, & \text{if } v(D,d) = v \text{ and } v \geq v_{esc}\\
        0,                           & \text{otherwise}
    \end{cases} \nonumber \\
    &=& C(D,V) \, p_e \, d^{-p_e-1} \, \delta(v - v(D,d)) \, H(v_e(D,d) - v_{esc}) \label{eqn.nEscape}
\end{eqnarray}
\noindent where $v(D,d)$ is expressed in \eqn{eqn.Hirase}, $\delta$ is the Dirac delta function, and $H$ is the Heaviside function.  

A $1\km$-diameter impactor at $\VmodeAtDonekm\kps$ excavates a crater of $\sim\oneKmImpactorCraterDiameterKM\km$ diameter with a volume of $\sim\oneKmImpactorCraterVolumeKMcubed\km^3$ and generates $\sim\oneKmImpactorEjectaGToneMeterInMillions\times 10^6$ ejecta larger than $1\meter$ diameter at greater than the lunar escape speed.

\subsection{The differential and cumulative lunar minimoon steady-state SFD}
\label{ss.SteadyStateSFD}

We can simplify \eqn{eqn.nSteadyState-full} by integrating over $v$, employing the $\delta$ and Heaviside functions in \eqn{eqn.nEscape}, and restricting the integral over $D$ to obtain
\begin{equation}
    n(d) \; \Delta d 
      = p_e \, d^{-p_e-1} \; \Delta d
         \int_{D_{min}} \dif D \int \dif V \; F'(D,V) \; C(D,V) \; \bar f(v(D,d)) \; \bar\ell(v(D,d)).
    \label{eqn.nSteadyStateFinal}
\end{equation}                       
\noindent where $v(D,d)$ is expressed in \eqn{eqn.Hirase} and the minimum diameter on the integral over the impactor diameter implements the Heaviside function.  We use $D_{min}=\DminMeters\meter$ as the minimum size impactor that can launch $1\meter$ diameter ejecta at the lunar escape speed.

The cumulative steady state number of minimoons, $N(d_{min})$, with $d \ge d_{min}$ is then
\begin{equation}
    N(d_{min}) 
    = p_e \, \int_{d_{min}}^\infty \dif d \; d^{-p_e-1}
              \int_{D_{min}} \dif D \int \dif V \; F'(D,V) \; C(D,V) \; \bar f(v(D,d)) \; \bar\ell(v(D,d)).
    \label{eqn.NSteadyStateFinal}
\end{equation}

\subsection{Lunar ejecta integrations}
\label{ss.LunarEjectaIntegrations}

In this sub-section we describe the dynamical integrations of synthetic particles ejected from the Moon's surface that will be used in \S\ref{ss.captureFraction} and \S\ref{ss.captureLifetime} to determine the final two terms in the integrand in equation \ref{eqn.nSteadyState-full}; $\bar f(v)$, the average fraction of lunar ejecta that become TBOs, and $\bar \ell(v)$, their average lifetime, both as a function of their ejection speed from the lunar surface which, in turn, depend on both the impactor's and ejecta's diameters, $v(D,d)$. 

\begin{enumerate}[itemsep=5pt]

    \item The simulation begins at an arbitrary time set to 1/1/2015~00:00~UT with launch epochs evenly distributed throughout one Metonic cycle of almost exactly 19 years, the period at which the relative positions of the Sun, Earth, and Moon nearly repeat in an inertial coordinate system.  The other massive bodies in the solar system will not have the same positions but these three bodies represent the dominant bodies under consideration.

    \item Each launch epoch corresponds to the formation of one of the \nCraters\ lunar craters. The position of the crater on the lunar surface, $\vec{r}_l=(x_l,y_l,z_l)$, is defined relative to the selenocentric coordinate system with the Moon's equatorial plane as the reference plane. The launch locations are randomly distributed over the lunar surface in latitude ($\phi$) and longitude ($\lambda$) and we used $R_M=1737.53\km$ for the radius of the Moon\footnote{This value is larger by $130\meter$ than the volumetric mean radius on the NASA fact sheet (\url{https://nssdc.gsfc.nasa.gov/planetary/factsheet/moonfact.html}) but 1) the notes on the fact sheet state that there are `no single set of agreed upon values' and 2) the values disagree by only $0.007$\%} \citep{Bills1977-LunarTopography}.
    
    \item At each launch location we assumed that all the particles were ejected at a fixed elevation angle of 45$\arcdeg$ with respect to the normal at the local surface.  Actual ejection angles vary as a function of impact angle, ejection location, impactor and surface properties, and other factors, but assuming a 45$\arcdeg$ ejection angle value is common, representative \citep[\eg][]{Cintala1999-impactEjecta,Anderson2003-ejectaFlow,Anderson2004-ejectionAngles}, and consistent with the measured cone angle from the DART impact mission \citep{Deshapriya2023-LICIACube-DARTEjectaPlume}.
    
    \item The ejecta's azimuth angles ($\beta$) were evenly distributed in the range $[0\mathopen{:}2\pi)$ with $\beta_j= 2\pi j/\nEjectaPerSpeed$ for $j=0,5$. 
    
    \item At each azimuth we launched \nSpeeds\ particles with a range of launch speeds in $\kps$ given by 
    \label{label.ejectionSpeeds}
    \begin{align*}
        \begin{aligned}
            v_0 & = \vZeroINkps\kps          &                   &     \\
            v_k & = v_0  + 0.01 \times 2^{k} & \hspace{20pt} k = & 0,6 \\
            v_m & = 3.02 + 0.28 \times m     & \hspace{20pt} m = & 1,5 \\
            v_n & = 4.42 + 0.14 \times n     & \hspace{20pt} n = & 1,7
        \end{aligned}
    \end{align*}
    where the minimum speed, $v_0$, is only $4\mps$ faster than the classical lunar escape speed.  \ie\ at each azimuth there are \nSpeeds\ fixed speeds of $v_0$, the 7 values of $k$, the 5 values of $m$, and 7 values of $n$.  The values were empirically chosen to span the range of speeds from barely escaping the lunar surface to having zero probability of becoming bound in the EMS.  The launch velocity with respect to the Moon's topocentric horizon at the launch location is then represented by ${\vec{v}_l}^{\; t}$. 

    \item We transform the topocentric velocity vector to the velocity vector $\vec{v}_l$ in the selenocentric lunar equatorial reference system following \citet{Bate1971-FundamentalsOfAstrodynamics}:
    \begin{equation}
        \vec{v}_l = \mathcal{R}_{3} ( \theta_{LST}) \; \mathcal{R}_2 ( 90\arcdeg - \phi ) \; {\vec{v}_l}^{\; t} + \vec{v}_h,
    \end{equation}
    where $\mathcal{R}_2$ and $\mathcal{R}_3$ are three-dimensional rotation matrices about the $y-$ and $z-$axis, respectively, $\theta_{LST}$ represents the local sidereal time (LST) on the Moon given by
    \begin{equation}
        \theta_{LST} = \lambda + \theta_0 + \omega_M ( t - t_0 ),
    \end{equation}
    where $\theta_0=70.79\arcdeg$ is the lunar LST at the reference time $t_0=$ 2023-Jun-04~03:15~UT from  JPL HORIZONS\footnote{\url{https://ssd-api.jpl.nasa.gov/doc/horizons.htm}\label{JPLHORIZONS}}, and $\vec{v}_h$ is the speed of the topocentric point due to the Moon's rotation around its axis,
    \begin{equation}
        \vec{v}_h = ( \; - \omega_M \; y_l, \omega_M \; x_l ,0 \; ),
    \end{equation}
    where $\omega_M \approx 2.66169 \times 10^{-6} \rad \second^{-1}$ is the Moon's rotation rate.

    \item We then determine the ejecta's solar system barycentric state vector
    \begin{equation}
        \vec{X} = ( \vec{r}, \vec{v} ) = \vec{X}_M + \mathcal{R}_1 ( \epsilon_l ) \; \vec{X}_l,
    \end{equation}
    where $\vec{X}_M$ is the lunacentric state vector at the launch epoch in the ecliptic reference system centered at the solar system barycenter obtained from  JPL HORIZONS$^{\ref{JPLHORIZONS}}$,
    $\vec{X}_l$ is the ejecta's lunacentric state vector, $\mathcal{R}_1$ is the rotation matrix about the $x-$axis, and $\epsilon_l = 1.54\arcdeg$ is the obliquity of the Moon relative to the ecliptic.
        
    \item Each particle's initial state vector $\vec{X}$ is propagated from its launch time using REBOUND \citep{ReinLiu2012} with a dynamical model that includes the Earth, Moon, Sun, and the other planets (a total of \nMassiveBodies\ massive bodies) whose state vectors at the launch time were obtained from the JPL ephemerides. The maximum integration time is \tIntegrationDays, about \tIntegrationYears.

    \item The propagation of every particle continues until 1) they collide with one of the massive bodies or 2) they leave the `extended intermediate source region' (xISR) region defined by heliocentric semi-major axes in the range $a_h\in[0.8\au,1.2\au]$, eccentricities in the range $e_h\in[0,0.2]$, and inclinations in the range $i_h\in[0\arcdeg,5\arcdeg]$ where the `h' subscript indicates heliocentric orbital elements.  The xISR is a superset of the orbital elements in the `intermediate source region' (ISR), the heliocentric orbital element phase space from which minimoons can be captured \citep{Fedorets2017-minimoons}:  $a_h\in[0.87\au,1.15\au]$, $e_h\in[0,0.12]$, and $i_h\in[0\arcdeg,2.5\arcdeg]$.
        
    \item We store a particle's state vector and orbital elements every $\minimoonSaveRateDays\Days$ when it has negative orbital energy with respect to the geocenter, otherwise every $\heliocentricSaveRateDays\Days$.
            
\end{enumerate}

\section{Results and Discussion}
\label{s.ResultsAndDiscussion}

\subsection{The average fraction of ejecta that become temporarily bound as a function of the ejecta's launch speed, $\bar f(v)$.}
\label{ss.captureFraction}

In the remainder of this work we make a distinction between `prompt' and `delayed' TBOs.  `Prompt' TBOs are those where the ejected particle becomes a TBO effectively immediately after ejection from the lunar surface while `delayed' TBOs are those that take some time to become TBOs, usually escaping from the EM system and orbiting the Sun for years to millions of years before being bound once again.  The distinction between a prompt and delayed TBO was empirically determined by examining the distribution of the time of the beginning of bound state, the moment at which the particles' orbital energy becomes negative with respect to the geocenter.  We found that there is an exponential decrease in the number of particles entering the TBO state as a function of time after ejection.  The rate levels out at about \MaxPromptStartDay~days so we consider any particle that becomes a TBO less than \MaxPromptStartDay~days after ejection as a prompt TBO.

\begin{figure}[htbp]
    \centering
    \includegraphics[width=0.48\columnwidth]{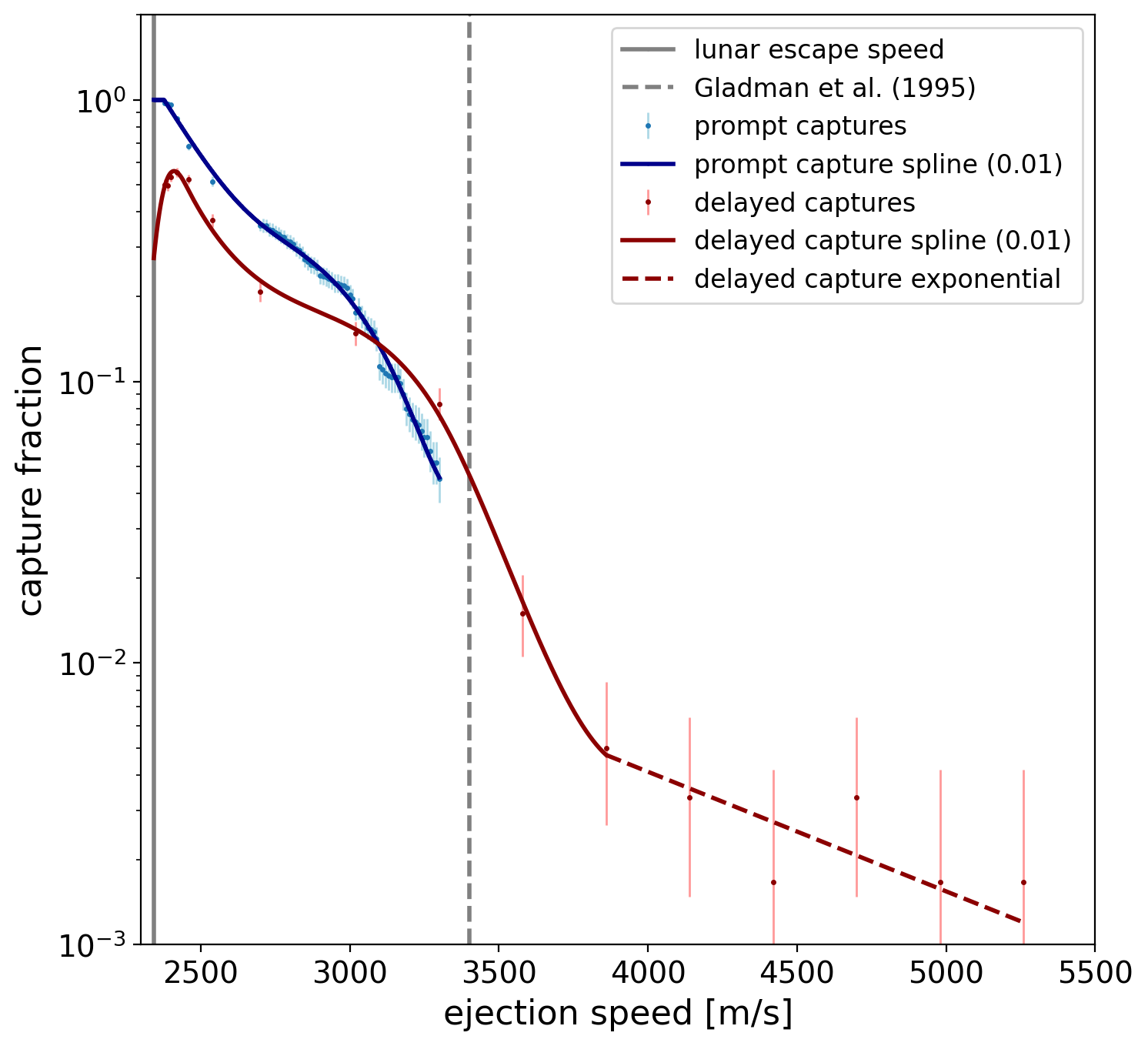}
    \includegraphics[width=0.48\columnwidth]{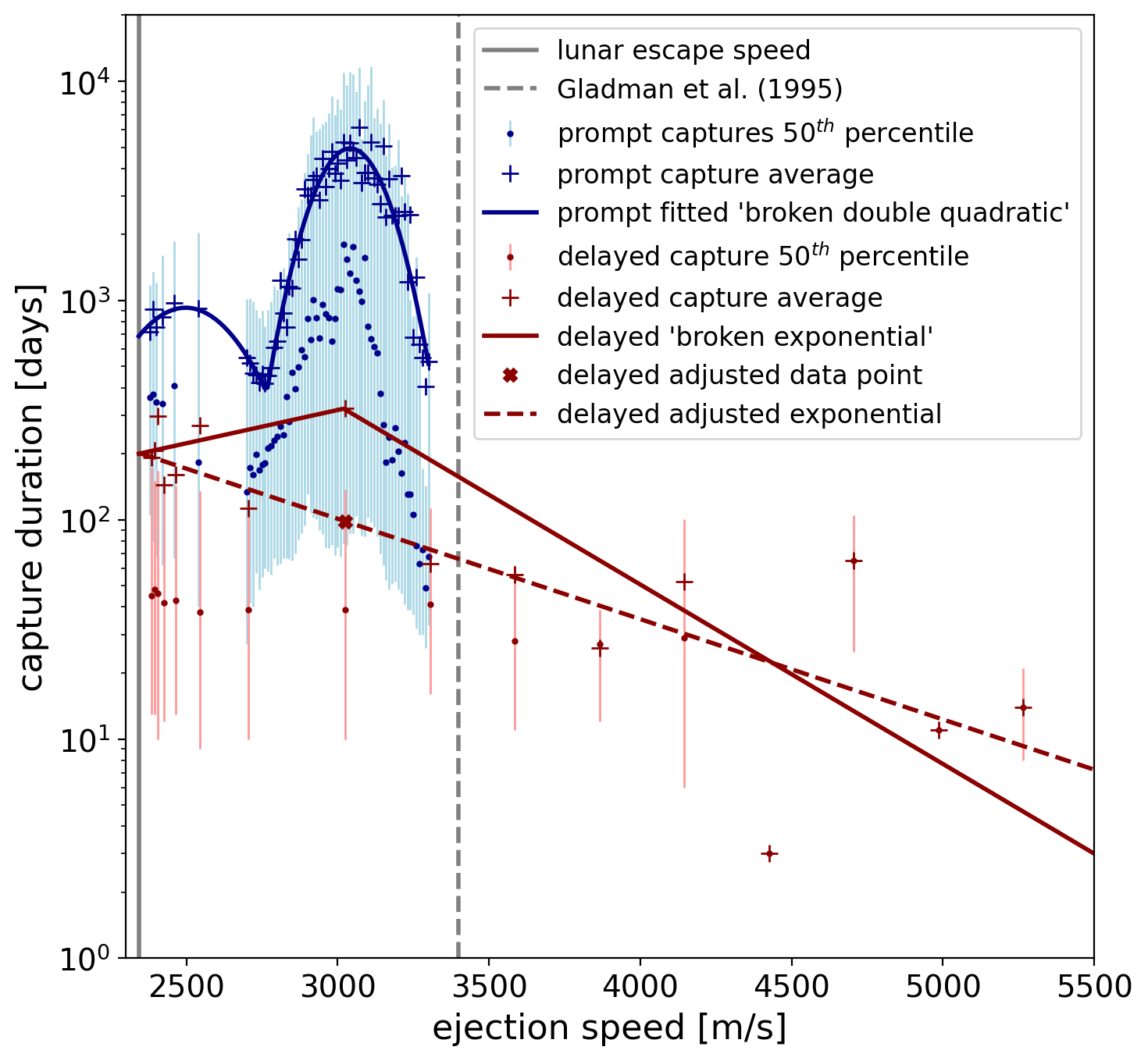}
    \caption{
        The fraction (left) and capture duration (right) of ejecta that become prompt (blue) and delayed (red) TBOs as a function of ejection speed.  
        The solid and dashed curves are fits to the data as explained in \S\ref{ss.captureFraction} and \S\ref{ss.captureLifetime}.
        In both panels the vertical solid grey line indicates the escape speed to leave the Moon as described in \S\ref{ss.n_ejecta} and the vertical dashed grey line indicates the ejection speed at which particles can not be bound in the EMS \citep{Gladman+.1995.Icarus-LunarImpactEjecta}.  The dotted data points and associated error bars in the right panel represent the median and the 16\ith\ and 84\ith\ percentile values in the cumulative distribution of duration times at each ejection speed.  The '+' data points represent the average duration at each ejection speed while the single `$\times$' data point at $3020\mps$ is the `adjusted' average capture lifetime at that speed obtained by removing just two exceedingly long-lived particles from the distribution. All the delayed data points are offset by $+\delayedSpeedPlotOffsetINmps\mps$ so as to not overlap the prompt data points.
        There were only a single delay-captured particle at ejection speeds of $4420\mps$ and $4980\mps$.}
    \label{fig.captureFraction+Lifetime}
\end{figure}

The fraction of ejecta that become TBOs as a function of their ejection speed was determined using the results of the integrations described in \S\ref{ss.LunarEjectaIntegrations}.  In general, the fraction of prompt and delayed TBOs decreases rapidly with a particle's ejection speed (\fig{fig.captureFraction+Lifetime}, left) with the only exception being that the fraction of delayed TBOs increases with ejection speed just above the lunar escape speed, presumably because most of the barely-escaping objects promptly become bound.  Effectively all the ejecta with launch speeds barely above the lunar escape speed are promptly bound while the maximum fraction for delayed TBOs is $\sim\maxDelayedCaptureFraction$.  A small fraction of ejecta with higher launch speeds can be bound long after ejection, presumably because gravitational perturbations of all the other massive objects in the integration modify their heliocentric orbits so that they can eventually become TBOs in the EM system.
 
We were unable to identify simple analytical expressions that could describe the shape of the prompt and delayed TBO fractions so we resorted to using \NumPy's \UnivariateSpline\ function with a smoothing factor of $s=\sCaptureFractionSpline$ and characterized the long tail on the delayed TBO fraction at high ejection speeds with a simple exponential function (\fig{fig.captureFraction+Lifetime}).

\subsection{Minimoons and temporary captures as a subset of the TBO population.}
\label{ss.MinimoonSubset}

As stated above, \citet{Granvik2012-minimoons} and \citet{Fedorets2017-minimoons} defined Earth's minimoons as TBOs that also make at least one revolution around Earth and pass within one Hill radius while a TBO. They defined temporarily-captured flybys (TCF) as TBOs that pass within one Hill radius without completing one revolution around Earth.

We determined that about 64\% of the delayed TBOs meet the TCF condition and about \TBOMinimoonPercentDelayed\% of the delayed TBOs meet the minimoon condition, almost independent of the particles' ejection speed from the lunar surface.

The situation is considerably different for the prompt TBOs, as expected.  About \TBOMinimoonRatioPercentLowSpeed\% of objects launched within $\sim60\mps$ of the lunar escape speed become prompt minimoons while only about \TBOMinimoonRatioPercentHighSpeed\% of the objects launched at speeds in the range $2700\mps$ to $3300\mps$, just less than the $3400\mps$ at which objects must escape the EMS \citep{Gladman+.1995.Icarus-LunarImpactEjecta}, become prompt minimoons.

\subsection{The average lifetime of ejecta that become TBOs as a function of the ejecta's launch speed, $\bar\ell(v)$.}
\label{ss.captureLifetime}

We determined the average lifetime of the TBOs, the time spent with $E_T<0$ relative to the geocenter, as a function of their ejection speed using the results of the integrations described in \S\ref{ss.LunarEjectaIntegrations}.  \ie\ Letting $\ell_i(v_j)$ represent the `lifetime' of the $i^{th}$ TBO that was ejected at speed $v_j$, the average lifetime of TBOs that were ejected at speed $v_j$ is
\begin{equation}
    {\bar\ell}(v_j) = \frac{1}{n_{mm}(v_j)} \sum_{i=1}^{n_{mm}(v_j)} \ell_i(v_j)
    \label{eqn.Lv}
\end{equation}
\noindent where $n_{mm}(v_j)$ is the total number of unique TBOs at the ejection speed.  We designate the particles' lifetimes as prompt and delayed TBOs as ${\bar\ell}_p(v_j)$ and ${\bar\ell}_d(v_j)$, respectively, and use the delayed capture lifetime for the steady-state calculation \ie\ $\bar\ell(v) \equiv \bar\ell_d(v)$.

Successive bound states of the same particle with a small time interval between them were merged before calculating the lifetimes. This occurs when the particles are far from Earth and their total energy with respect to the geocenter is briefly $>0$ due to perturbations in the integration's $n$-body implementation. The most likely time separation between successive bound states is about 1~day with a tail dropping off rapidly for longer intervals and reaching a minimum at about \maxTimeBetweenSuccessiveCapturesDays~days.  Thus, we chose \maxTimeBetweenSuccessiveCapturesDays~days as the minimum interval between independent successive TBO states. \ie\ a TBO that briefly becomes unbound for $<\maxTimeBetweenSuccessiveCapturesDays$~days and then enters another TBO phase is considered a single TBO event.

The bound durations of particles at each ejection speed had long tails so we present the data (\fig{fig.captureFraction+Lifetime}) with `error' bars around the median values representing the 16\ith\ and 84\ith\ percentiles in the lifetimes at each speed with the intent that those percentiles are proxies for `1-sigma' uncertainties with respect to the median values.

Using the set of 20 coarsely-separated ejection speeds defined by item \ref{label.ejectionSpeeds} in \S\ref{ss.LunarEjectaIntegrations} we were surprised to identify anomalously high prompt TBO durations for particles ejected at $3020\mps$, so we integrated many more particles with ejection speeds ranging from $2700\mps$ to $3300\mps$ in steps of $10\mps$ for $10^5\Days$.  These finely-separated integrations confirmed that there is a broad range of ejection speeds for which particles promptly became TBOs with long durations (\fig{fig.captureFraction+Lifetime}, right) beginning at speeds of about $\promptQuadraticBreakSpeedMPS\mps$, peaking at $\sim\maxAvgPromptCaptureEjectionSpeedMPS\mps$ and then falling rapidly as the ejecta's speed approaches $3400\mps$, the launch speed at which particles must escape the lunar sphere of influence \citep{Gladman+.1995.Icarus-LunarImpactEjecta}.  The long-duration trajectories are mainly retrograde with respect to the Earth and prefer apocenters near the edge of the Earth's sphere of influence. The Moon seems to have a secondary role in their long duration. A deeper analysis will be provided in a subsequent paper.

Our prompt TBO durations are in good agreement with previous work \citep[\eg][]{Gault1983-TerrestrialAccretionofLunarMaterial,Gladman+.1995.Icarus-LunarImpactEjecta}.  \citet{Gladman+.1995.Icarus-LunarImpactEjecta} performed similar 4-body integrations with more than 8000 particles for a period of 300 years and found that none survived in geocentric orbit for the entire time.  Our \nMassiveBodies-body integrations of  12,000 particles also have no prompt TBOs surviving for three centuries, 99\% of them have lifetimes of $<\promptDurationNinetyNinePercentileYears\yr$, and 90\% have bound durations $<\promptDurationNinetyPercentileYears\yr$.

The delayed TBO behavior is explicable, with generally decreasing average durations as the ejection speed increases except for the anomalously long lifetime at, once again, $3020\mps$.  A total of 148 particles evolved into delayed TBOs at that launch speed and two of the captures were in the EMS for more than $10,000\Days$, almost 30 years.  Ignoring those two particles reduces the average TBO duration to about $100\Days$, perfectly in line with expectations based on TBO durations at the other speeds (\fig{fig.captureFraction+Lifetime}, right).

Once again, we were unable to identify simple analytic expressions to fit the prompt and delayed TBOs, and \NumPy's \UnivariateSpline\ function could not provide satisfactory representations of the data, so we resorted to piecewise continuous functions.  

We fit a `broken double quadratic' function in $\log({\bar\ell}_p(v_j)/\mathrm{days})$ vs. $v_j/\mathrm{\mps}$ to the data and required that the two quadratics have the same values at their intersection point (\fig{fig.captureFraction+Lifetime}, right).  A piecewise continuous linear function was empirically defined to represent the delayed TBO durations in $\log({\bar\ell}_d(v_j)/\mathrm{days})$ vs. $v_j/\mathrm{\mps}$ such that the `break point' is at $3020\mps$ and the minimum speed is the lunar escape speed (\fig{fig.captureFraction+Lifetime}, right).  Finally, after removing the two particles with anomalously long capture durations at $3020\mps$, we were able to fit a simple exponential function to the average delayed TBO durations (\fig{fig.captureFraction+Lifetime}, right).

\subsection{Steady state lunar-source TBO and minimoon size-frequency distribution}
\label{ss.minimoonSFD}

\begin{figure}[!ht]
    \centering
    \begin{minipage}{10cm}  
        \includegraphics[width=\columnwidth]{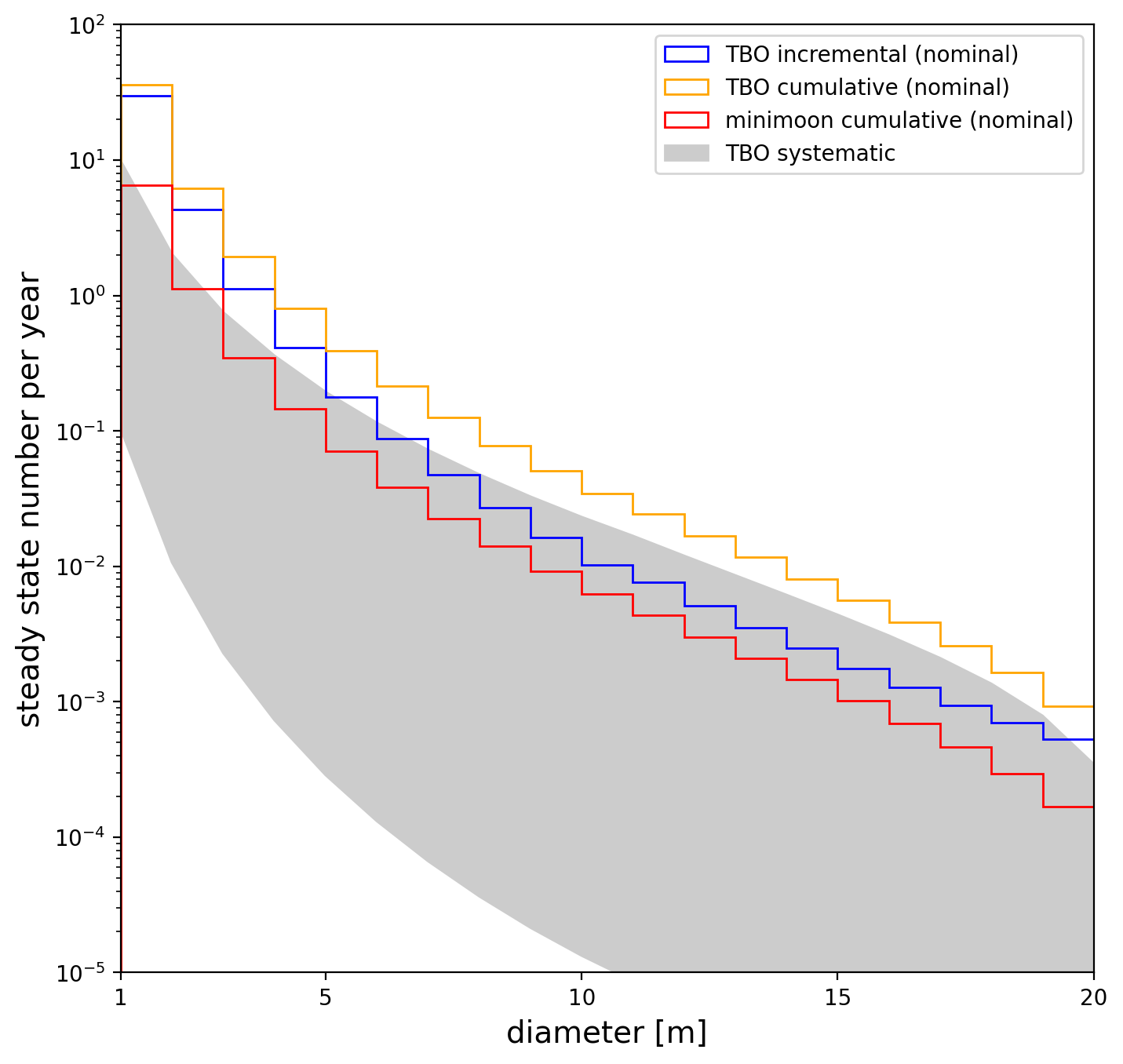}
        \caption{
            The nominal incremental (blue) and cumulative (orange) steady state annual number of lunar-impact-generated TBOs larger than $1\meter$ diameter. The nominal cumulative SFD of minimoons (red) is \TBOMinimoonPercentDelayed\% of the TBO distribution as explained in \S\ref{ss.MinimoonSubset}. The gray band represents the range of possible cumulative distributions given the uncertainties in our input distributions when we restrict the cumulative number of TBOs $>1\meter$ in diameter to between 0.1 and 10 as described in \S\ref{ss.minimoonSFD}.}
        \label{fig.SteadyStateMinimoonSFD}
    \end{minipage}
\end{figure}

At this point we have functions representing all the factors in \eqn{eqn.nSteadyState-full} and can perform the integration to calculate the steady state annual number of lunar-impact-generated TBOs larger than $1\meter$ diameter (\fig{fig.SteadyStateMinimoonSFD}).  Our nominal prediction is that there are $\nMinimoonsOneMeterCumulative$ TBOs larger than $1\meter$ in diameter of lunar origin per year in the EMS and $\nMinimoonsOneMeterIncremental$ of those TBOs are in the $1\meter$ to $2\meter$ diameter range due mostly to the steep size-frequency distribution of the ejecta (\S\ref{ss.n_ejecta}).  Given that \TBOMinimoonPercentDelayed\% of TBOs can also be classified as minimoons (\S\ref{ss.MinimoonSubset}) our nominal results suggest that there should be about $6.5$ minimoons larger than $1\meter$ diameter in the EMS at any time.

With the results of our lunar ejecta integrations in hand (\S\ref{ss.LunarEjectaIntegrations}) we can address whether our assumption that the population can be treated as a `steady state' is appropriate.  The time scale for lunar impacts that can generate objects that can become TBOs is about once per 40,000 years (\fig{fig.ImpactFlux+SpeedDistn}) while the average time of the beginning of a TBO state is $\sim75,000$ years.  Large but infrequent impactors will inject a significant pulse of lunar ejecta into the xISR while the smaller, more frequent, impactors supply a `steady' stream of objects.  Thus, our calculation may be considered a `long term average' if the `steady state' term is unpalatable. 

Our nominal value is about $5\times$ larger than \citet{Granvik2012-minimoons}'s calculation that there is a single minimoon larger than $1\meter$ in diameter captured from the NEO population at any time, which we convert to an annual number of $\sim1.3$ minimoons given that their reported average minimoon lifetime is about 286~days.  \citet{Fedorets2017-minimoons}'s more detailed analysis of the population came to the conclusion that there are about $0.8$ minimoons larger than $1\meter$ diameter at any time.  They included a more sophisticated NEO orbital distribution model and distinguished between `orbiters' and `flybys'.  The weighted average lifetime of the population of minimoons is then about $153\days$
so the annualized number of minimoons is about 1.9, $\gtrsim3\times$ smaller than our nominal value of TBOs.

The actual number of minimoons $>1\meter$ in diameter at any time is unknown because of the difficulties in accounting for observational selection effects for these small, fast-moving objects.  Two minimoons in this size range have been discovered in the past twenty years, \RH\ and \CD, so the steady state annual number is probably $>0.1$.  Contemporary asteroid surveys regularly detect geocentric objects on minimoon-like orbits but almost all of them turn out to be unusual distant artificial objects \citep[\eg,][]{Battle202-2020SO}.  Given the capabilities of, and our experience with, modern asteroid surveys, we suggest that the annual number of minimoons larger than 1~meter in diameter is probably $<10$.  If the annual number of minimoons was 10 and they followed the \citet{Dohnanyi1969} cumulative size distribution $N(>d) \propto d^{-p}$ with $p=2.5$, then we would expect there to be one $5\meter$ diameter or larger minimoon about every 5 or 6 years and a natural object of that size would almost certainly be discovered by one of the surveys.  If the minimoon SFD was steeper, with $p=\bar{p_e}=3.7$ (\S\ref{ss.n_ejecta}), then minimoons with $d>5\meter$ would be much less common, occurring just once every four decades, but that would still imply an $\sim50$\% chance of such an object in the past twenty years.  Thus, our nominal lunar-impact-generated TBO population is at least a few times larger than the actual population and probably an order of magnitude or more.  Furthermore, since \citet{Granvik2012-minimoons} established that minimoons can derive through orbital evolution from the MB, the actual population of minimoons could be entirely explained with the MB as the parent population rather than lunar impacts.

\begin{table}[]
    {\renewcommand{\arraystretch}{1.25}  
    \centering
    \begin{tabular}{l|l|p{11cm}}
                                           & \textbf{term} & \textbf{definition}  \\ \hline\hline
        \multirow{4}{*}{\rotatebox{90}{\textbf{impactor}}} & $D$           & diameter                                                                                                       \\ \cline{2-3} 
                                           & $V$           & speed                                                                                                          \\ \cline{2-3} 
                                           & $F(D)$        & flux density at diameter $D$                                                                                   \\ \cline{2-3} 
                                           & $p(D,V)$      & probability density of speed $V$ at diameter $D$                                                               \\ \hline\hline
        \multirow{7}{*}{\rotatebox{90}{\textbf{ejecta}}}   & $n(d)$        & steady-state number of TBOs as a function of diameter $d$                                                      \\ \cline{2-3} 
                                           & $d$           & diameter                                                                                                       \\ \cline{2-3} 
                                           & $v$           & speed                                                                                                          \\ \cline{2-3} 
                                           & $n_{escape}$  & number density of escaping ejecta at diameter $d$ and speed $v$ given an impactor of diameter $D$ at speed $V$ \\ \cline{2-3} 
                                           & $\bar f(v)$   & average fraction of ejecta with speed $v$ that ever become TBOs                                                \\ \cline{2-3} 
                                           & $\bar\ell(v)$ & average lifetime as TBOs of ejecta with speed $v$                                                              \\ \hline\hline
    \end{tabular}
    \caption{Definition of terms in \eqn{eqn.nSteadyState-full}.}
    \label{tab.eqn2termDefinitions}
    }  
\end{table}

\subsection{Steady state lunar TBO size-frequency distribution systematics}
\label{ss.minimoonSFD_systematics}

\newcommand{\rhoS}{{ \SdensitykgPerCubicMeter    \kg\meter^{-3}}}
\newcommand{\drhoS}{{\SdensitykgPerCubicMeterUnc \kg\meter^{-3}}}

\newcommand{\rhoC}{{ \CdensitykgPerCubicMeter    \kg\meter^{-3}}}
\newcommand{\drhoC}{{\CdensitykgPerCubicMeterUnc \kg\meter^{-3}}}

\begin{table}[]
    \centering
    \begin{tabular}{lcccl}
        \hline
        parameter/function              & symbol       & nominal                            & range                             & influence \\
                                        &              &  value                             &                                   &           \\ \hline
        average impact angle            & $\theta$     & $45\arcdeg$                        & $\pm25\arcdeg$                    & none      \\ \gline
        Earth:Moon impact ratio$^a$     & $R_{E:M}$    & \EarthMoonImpactRatio              & $\pm\EarthMoonImpactRatioUnc$     & none      \\ \gline
        S fraction$^b$                  & $f_S$        & \fLunarSImpactors                  & $\pm\fLunarSImpactorsUnc$         & minimal   \\ \gline
        S density$^c$                   & $\rho_S$     & $\rhoS$                            & $\pm\drhoS$                       & none      \\ \gline
        C density$^c$                   & $\rho_C$     & $\rhoC$                            & $\pm\drhoC$                       & none      \\ \gline
        crater depth:diameter ratio$^d$ & $\Re$        & \craterDepthToDiameter             & $\pm 0.02$                        & minimal   \\ \gline
        crater shape                    & n/a          & paraboloidal                       & paraboloidal                      & none      \\
                                        &              &                                    & hemispherical                     &           \\ \gline
        crater scaling relation         & n/a          & [3]                                & [1-5]                             & minimal   \\ \gline
        ejecta SFD exponent$^e$         & $p_e$        & \ejectaSFDslope                    & $\pm\ejectaSFDslopeRMS$           & major     \\ \hline
    \end{tabular}
    \caption{The set of parameters and functions that were randomly varied as described in \S\ref{ss.minimoonSFD_systematics} to study their effect on the predicted lunar TBO SFD.
                $^a$\citet{Nesvorny2024-NEOMOD2}, \citet{Werner2002-LunarImpactorSFD}
                $^b$\citet{Wright2016}
                $^c$\citet{Carry2012}
                $^d$\citet{Pike1974-LunarDepthDiameterRelations}
                $^e$\citet{Bart2010-BouldersEjectedFromLunarCraters}
                [1] \citet{Singer2020-EjectaBlockSizes}  
                [2] \citet{Horedt1984-SixCraterScalingLaws}
                [3] \citet{Collins2005-craterScaling}
                [4] \citet{Bottke2016-LPSC-CraterScalingLaws}
                [5] \citet{Housen+Holsapple2011-EjectaFromImpactCraters}.}
    \label{tab.SystematicStudy}
\end{table}

The input parameters for our nominal lunar TBO steady state annual SFD (\fig{fig.SteadyStateMinimoonSFD}) were selected based on our interpretation of representative values from the literature but yielded a result much larger than suggested by the observed population.  To characterize the effect of each of the parameters on our nominal result we implemented a systematic study of the impact of most of the important input parameters (\tab{tab.SystematicStudy}).  We randomly generated a set of new values for each of the parameters, or randomly selected a function, and then re-calculated the minimoon SFD.  The random numerical parameters were generated using a normal function with a mean set to the nominal value and sigma set to the range as defined in \tab{tab.SystematicStudy} but truncated at $\pm1\sigma$ so as not to introduce outrageous values.

We ran \nSystematics\ simulations and found that the predicted number of TBOs larger than $1\meter$ diameter ranged over \sysNOrders\ orders of magnitude, from \sysNMin\ to \sysNMax, with an average value of \sysNAvg, 1-2 orders of magnitude larger than the actual population, despite the truncation in the range of parameter values.  

We then restricted the analysis to the \nSystematicsGood\ `good' simulations that yielded between 0.1 and 10 TBOs $>1\meter$ diameter per year under the assumption that the entire TBO population derives from lunar impacts.  The fraction of simulations that were good is independent of the average impact angle, the Earth:Moon impact ratio, the densities of S- and C-class asteroids, and whether the craters have paraboloidal or hemispherical shapes.  The fraction slightly favors shallower crater depths and a higher fraction of S-class impactors in the population than the nominal value.

\begin{figure}[htbp]
    \centering
    \begin{minipage}{15cm}  
        \includegraphics[width=\linewidth]{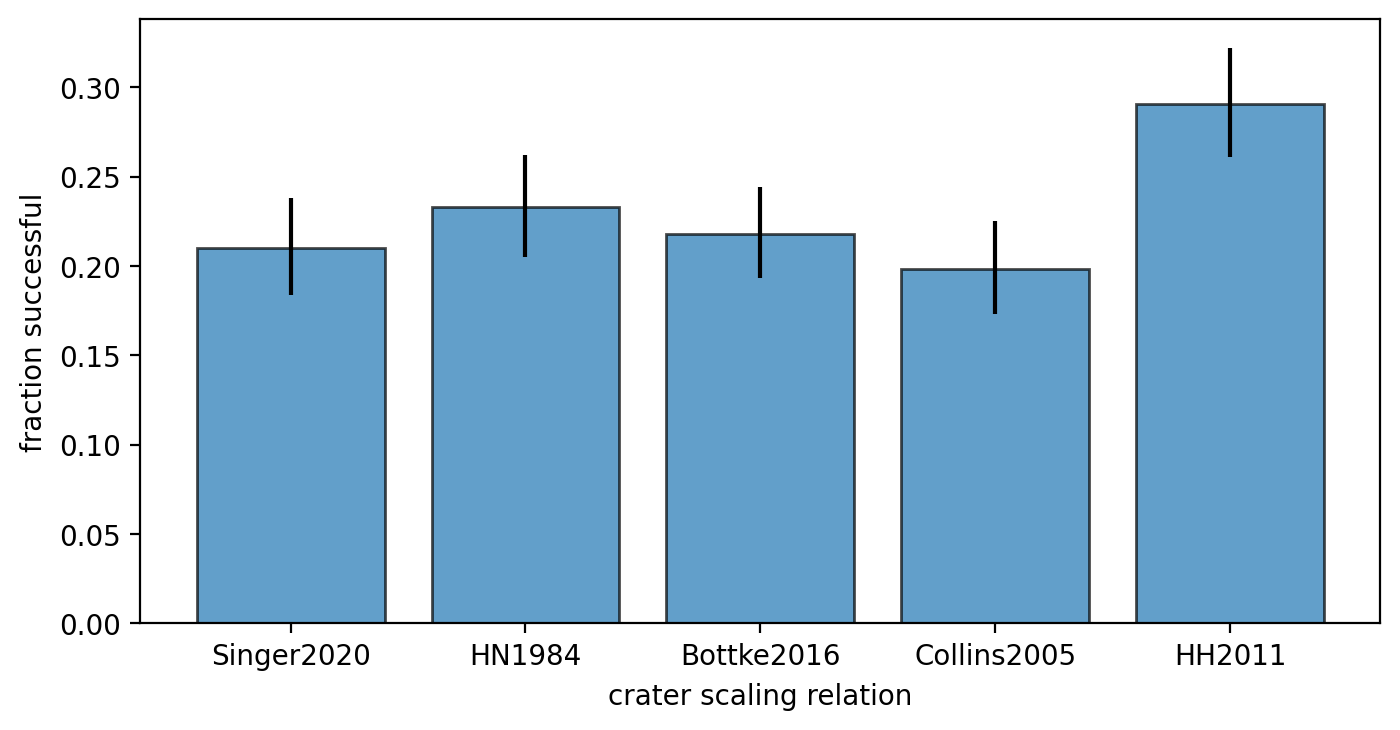}
        \caption{The success rate of the five crater scaling relations at generating between 0.1 and 10 TBOs $>1\meter$ diameter per year 
            in our systematic studies described in \S\ref{ss.minimoonSFD_systematics} \citep{Singer2020-EjectaBlockSizes,Horedt1984-SixCraterScalingLaws,Bottke2016-LPSC-CraterScalingLaws,Collins2005-craterScaling,Housen+Holsapple2011-EjectaFromImpactCraters}.}
        \label{fig.CraterScalingRelationsSuccessRate}
    \end{minipage}
\end{figure}

All five of the crater scaling functions (\fig{fig.CraterScalingRelations}) could generate results that yielded between 0.1 and 10 TBOs $>1\meter$ diameter per year but the \citet{Housen+Holsapple2011-EjectaFromImpactCraters} scaling relation was significantly better than our nominal relation \citep{Collins2005-craterScaling} and the other three (\fig{fig.CraterScalingRelationsSuccessRate}).  Given that the scaling relations differ by $\sim4\times$ in the predicted crater diameter at a given impactor diameter (\fig{fig.CraterScalingRelations}), implying about a couple orders of magnitude in the volume of excavated material at fixed impactor diameter, it is not surprising that the relations require different ejecta SFD to satisfy our condition on the actual number of TBOs (\tab{tab.CraterScalingSlopes}).  \citet{Singer2020-EjectaBlockSizes} and \citet{Horedt1984-SixCraterScalingLaws} produce the largest craters and therefore require the steepest ejecta SFD slopes with $p_e=\HNSlopeAvg\pm\HNSlopeStd$ while the \citet{Housen+Holsapple2011-EjectaFromImpactCraters} relation yields the smallest craters and requires a significantly shallower ejecta SFD with $\HHSlopeAvg \pm \HHSlopeStd$.  Our nominal scaling relation, \citet{Collins2005-craterScaling}, and the straightforward \citet{Bottke2016-LPSC-CraterScalingLaws} relation produce intermediate-sized craters with intermediate ejecta SFD but they are numerically equal to the \citet{Housen+Holsapple2011-EjectaFromImpactCraters} value.  All our ejecta SFD slopes are consistent with our nominal value of $p_e = \ejectaSFDslope \pm \ejectaSFDslopeRMS$.

The $\sim3$\% uncertainty on the ejecta SFDs in \tab{tab.CraterScalingSlopes} compared to the $\sim25$\% uncertainty on the slope determined by \citet{Bart2010-BouldersEjectedFromLunarCraters} using lunar boulders suggests that \emph{if} the number of TBOs/minimoons with lunar provenance can be established and, even better, if it is possible to determine their SFD, the TBO/minimoon population could be used to constrain crater scaling relations and the ejecta SFD.  Establishing whether TBOs/minimoons have a lunar provenance will probably be determined through spectroscopic observations.

We stress that we are \emph{not} claiming that we have measured the crater ejecta SFD to $\sim3$\% uncertainty or are promoting any of the crater scaling relations used in this work.  The steady state number of TBOs larger than $1\meter$ diameter is uncertain by perhaps two orders of magnitude and we do not know if any of them derive from the Moon.  We have only shown that lunar impacts \emph{could} generate a TBO population consistent with the actual population with an appropriate combination of a crater scaling relation, ejecta SFD, and ejecta size-speed relation.

\begin{table}[]
    \centering
    \begin{minipage}{10cm}  
        \begin{tabular}{l|c}
        \hline
             Crater Scaling Relation                                &    Ejecta SFD slope ($p_e$)                   \\
             \hline
             \citet{Singer2020-EjectaBlockSizes}                    &    $ \SingerSlopeAvg \pm \SingerSlopeStd$     \\
             \citet{Horedt1984-SixCraterScalingLaws}                &    $     \HNSlopeAvg \pm     \HNSlopeStd$     \\  
             \citet{Bottke2016-LPSC-CraterScalingLaws}              &    $ \BottkeSlopeAvg \pm \BottkeSlopeStd$     \\
             \citet{Collins2005-craterScaling}                      &    $\CollinsSlopeAvg \pm \CollinsSlopeStd$    \\
             \citet{Housen+Holsapple2011-EjectaFromImpactCraters}   &    $     \HHSlopeAvg \pm      \HHSlopeStd$    \\
             \hline
        \end{tabular}
        \caption{The mean and standard deviation of the slope of the ejecta SFD that allows the crater scaling relation to generate between 0.1 and 10 TBOs $>1\meter$ diameter per year, a value consistent with the actual number of TBOs.}
        \label{tab.CraterScalingSlopes}
    \end{minipage}
\end{table}

\subsection{Geocentric residence time distributions}
\label{ss.GeocentricResidenceTimeDistributions}

We define a geocentric orbital residence time distribution, $t_{R,k}(v,a,e,i;\Delta a,\Delta e, \Delta i)$, as the amount of time the $k$\ith\ TBO ejected from the lunar surface at speed $v$ has geocentric orbital elements in the simultaneous ranges $[a-\Delta a/2,a+\Delta a/2]$, $[e-\Delta e/2,e+\Delta e/2]$, and $[i-\Delta i/2,i+\Delta i/2]$:
\begin{equation}
    t_{R,k}(v,a,e,i;\Delta a,\Delta e, \Delta i) 
    = \int B[a_k(v,t);a,\Delta a] \;
           B[e_k(v,t);e,\Delta e] \;
           B[i_k(v,t);i,\Delta i] \; dt
\end{equation}
where the boxcar function is 
\begin{equation}
    B( x; y, \Delta y ) =
    \begin{cases}
        1 & \text{if } y - \Delta y/2 \leq x < y + \Delta y/2 \\
        0 & \text{otherwise}.
    \end{cases}
\end{equation}

The geocentric orbital residence time probability density, $p_R'(v,a,e,i)$, for lunar TBOs ejected from the Moon's surface at speed $v$ is defined as
\begin{equation}
    p_R'(v,a,e,i) 
    = \frac{1}{n_k \, \Delta a \, \Delta e \, \Delta i} 
      \sum_{k=1}^{n_k} \; t_{R,k}(v,a,e,i;\Delta a,\Delta e, \Delta i)
\end{equation}
and the summation occurs over the $n_k$ TBOs that were captured after being ejected at speed $v$.  

We then perform a weighted integral over the ejection speed to arrive at the average geocentric orbital residence time probability density, $p_R(a,e,i)$, for TBOs ejected from the lunar surface:
\begin{equation}
    p_R(a,e,i) = \frac{ \int n(v) \, \bar{f}(v) \; p_R'(v,a,e,i) \; \dif v}
                      { \int n(v) \, \bar{f}(v)                  \; \dif v},
    \label{eqn.pR}
\end{equation}
where $p_R'$ is weighted by the number of ejecta at the initial ejection speed $v$, $n(v)$, and the fraction of particles, $\bar{f}(v)$, that are bound which were originally launched at speed $v$ (\S\ref{ss.captureFraction}).  Following the derivation of \eqn{eqn.nSteadyStateFinal},
\begin{equation}
    n(v) = p_e \; \int_{D_{min}} \dif D \int \dif V \; F'(D,V) \; C(D,V) \; d(D,v)^{-p_e-1} \; H[d-\min(d)] \; \Big|_{v>v_{esc}},
    \label{eqn.nv}
\end{equation}
\noindent where $\min(d)$ is the minimum TBO diameter of $1\meter$, and we explicitly show the dependence of the ejecta diameter on the impactor's diameter and speed in our formalism (\eqn{eqn.Hirase}).

While $n(v)$ and $\bar{f}(v)$ are continuous functions we only dynamically integrated unique values, $v_j$, of the ejection speeds and $\bar{f}(v)$ is different for prompt and delayed TBOs (\S\ref{ss.LunarEjectaIntegrations}).  Thus, to evaluate \eqn{eqn.pR} we treat $p_R'(v,a,e,i)$ as a piecewise continuous function where
\begin{equation}
    p_R'(v,a,e,i) = p_R'(v_j,a,e,i) \quad \mathrm{for} \quad \frac{v_{j-1}+v_j}{2} <= v < \frac{v_j+v_{j+1}}{2}.
\end{equation}

\begin{figure}[htbp]
    \centering
    \includegraphics[width=0.48\columnwidth]{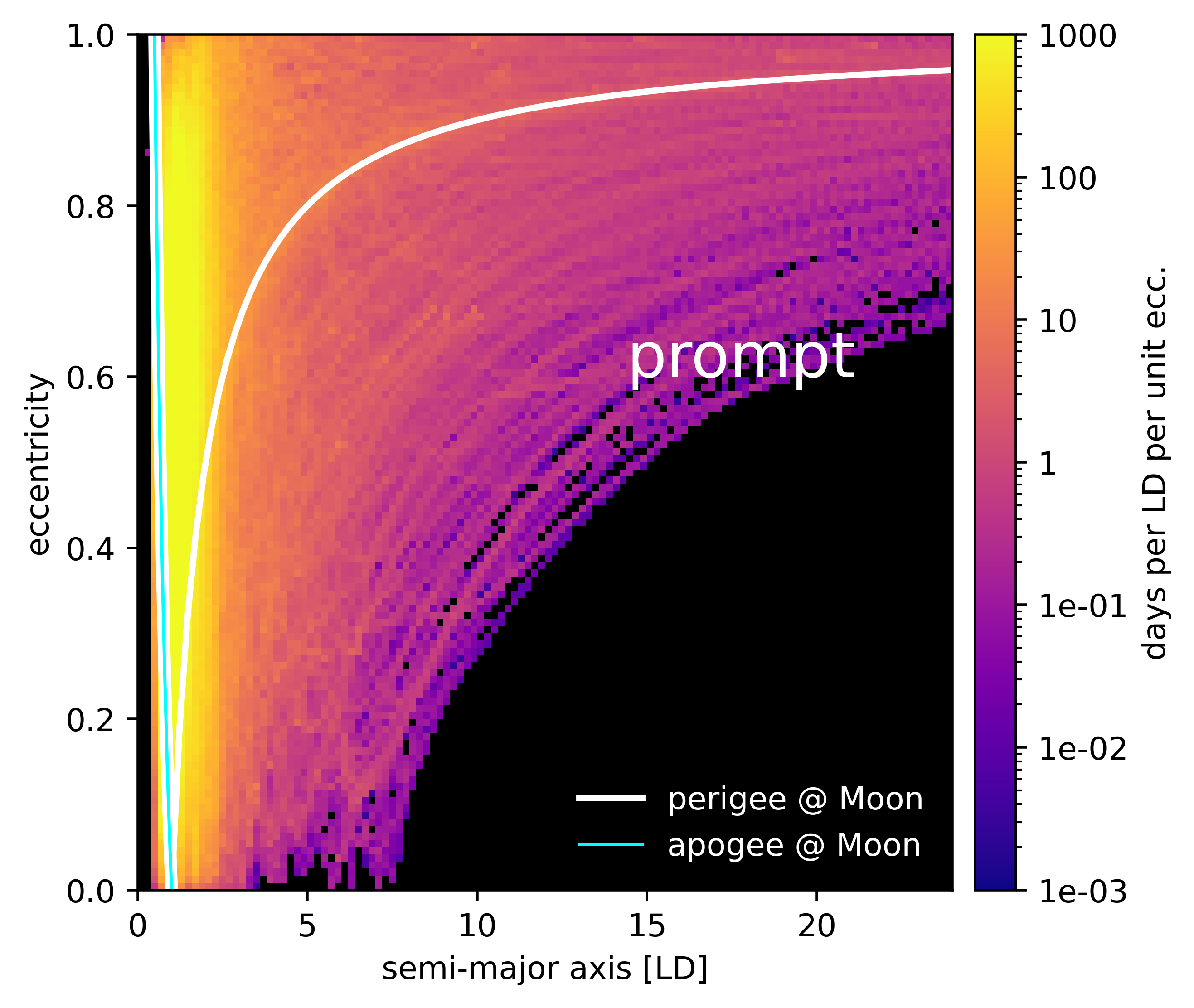}
    \includegraphics[width=0.48\columnwidth]{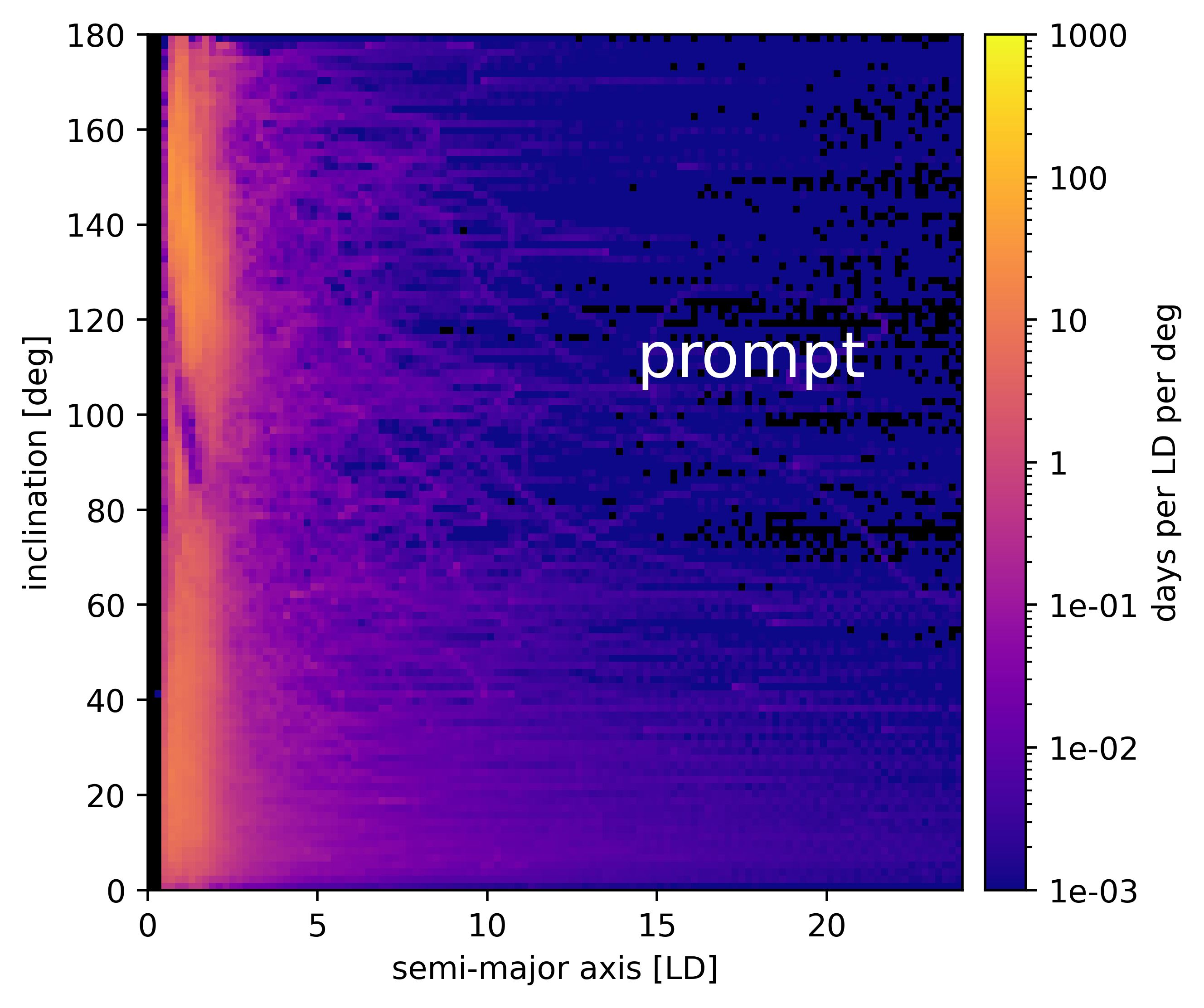}
    \includegraphics[width=0.48\columnwidth]{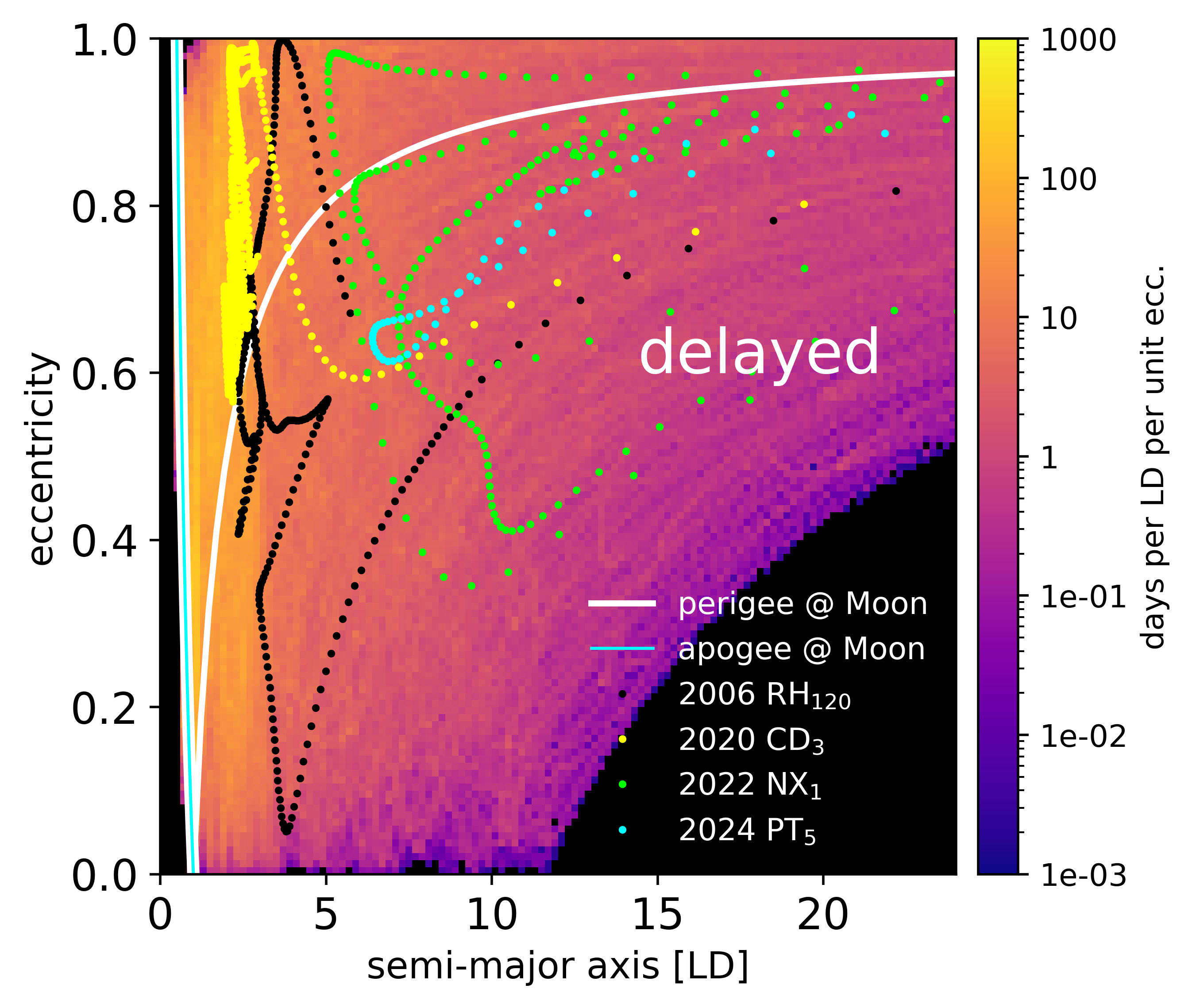}
    \includegraphics[width=0.48\columnwidth]{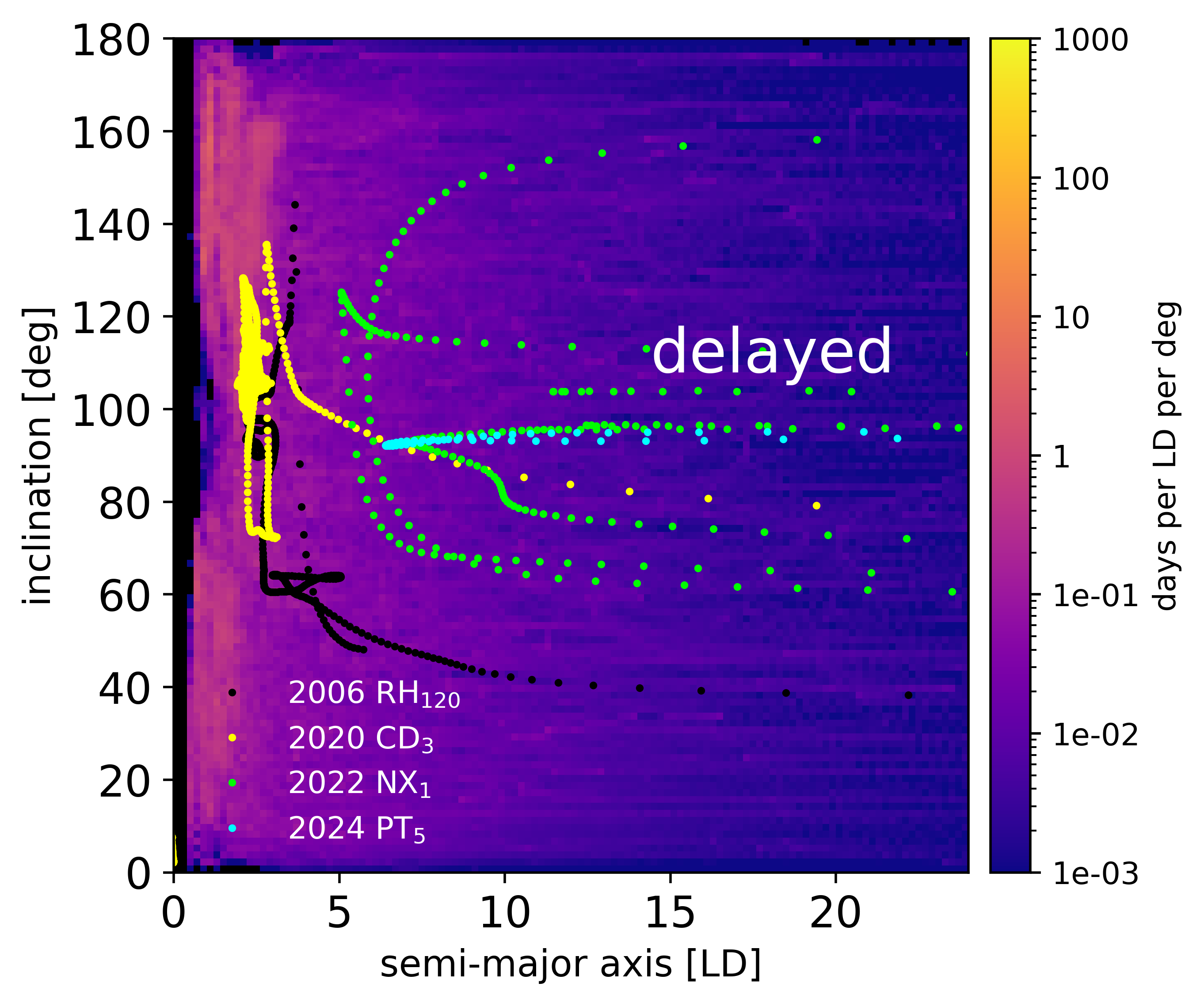}
    \caption{
        Residence time distributions for TBOs of lunar provenance as a function of
        (left) eccentricity vs. semi-major axis and
        (right) inclination vs. semi-major axis for
        (top) `prompt' TBOs, those that are bound in the EMS within 10 days after ejection from the lunar surface, and
        (bottom) 'delayed' TBOs, which become bound more than 10 days after ejection.
        The left panels show curves representing the orbital elements corresponding to the object being at the Moon's distance at apogee (cyan with white border) and perigee (white).
        The lower panels for the delayed TBOs provide the geocentric orbital elements on a daily basis of 4 known TBOs.  It includes all 5 periods of TBO status experienced by \NX\ from 1980 through 2052.}
    \label{fig.residenceTimeDensity}
\end{figure}

We calculate separate residence time probability densities for both prompt and delayed TBOs since the capture fractions and lifetimes are different for the two populations.  Since it is difficult to display a three-dimensional residence time we integrate over one of the orbital elements to result in a two dimensional residence time distribution \eg\ integration over the inclination to obtain $t_R(a,e;\Delta a,\Delta e)$ (\fig{fig.residenceTimeDensity}).

Both \citet{Granvik2012-minimoons} and \citet{Fedorets2017-minimoons} provided residence time distributions for their populations of natural Earth satellites (NES) composed of temporarily-captured orbiters (TCO), and the latter also provided the distributions for temporarily-captured flybys (TCF).  The source of their NESs was the population of small asteroids that have dynamically evolved out of the MB onto Earth-like orbits, a class of objects that must be dynamically similar to the lunar ejecta consider in this work that escape the EM system into heliocentric Earth-like orbits.  Thus, we expected that the residence time distribution of their NESs should be similar to our delayed captures and there are qualitative similarities between their work and ours including: a lack of objects with semi-major axis less than 1~LD, favoring objects with geocentric semi-major axis $a_g\lesssim3$~LD, and a semi-major axis limit of $\sim12$~LD for low-eccentricity objects caused by our identical requirement that the objects be within $3\;R_H$ of the EM system.

There are notable differences between the three sets of residence time distributions.  \citet{Granvik2012-minimoons} noted an island of long-lived NESs with semi-major axes $\sim0.5$~LD and $i\sim35\arcdeg$ that is not present in \citet{Fedorets2017-minimoons} or this work, and both of their results suggest that retrograde NESs are more common than prograde objects.

\begin{table}[]
    \centering
    \begin{minipage}{10cm}  
        \begin{tabular}{r|c|c}
            condition       & prompt    & delayed \\
            \hline
            inside Earth's Hill sphere  &  99\%  &   78\%  \\
outside Earth's Hill sphere  &   1\%  &  22\%  \\
prograde ($i<90\arcdeg$)  &  34\%  &           38\%  \\
retrograde ($i\ge90\arcdeg$)  &  66\%  &         62\%  \\
low ecc. ($e<0.5$) w/in 2 R$_H$  &  43\%  &               37\% \\
high ecc. ($e\ge0.5$) w/in 2 R$_H$  &  57\%  &              58\%  \\

        \end{tabular}
        \caption{Percentage of time that prompt and delayed TBOs with lunar provenance meet the stated condition.  Values may not add to 100\% because the orbital element phase space may not be fully covered.}
        \label{tab:residenceTimePercentages}
    \end{minipage}
\end{table}

While \fig{fig.residenceTimeDensity}\ illustrates the orbital element distribution for TBOs of lunar provenance it provides a misleading representation of their time in the phase space due the logarithmic scale in the residence time.  The prompt TBOs of lunar provenance spend almost all their time inside Earth's Hill sphere on prograde orbits  and the delayed captures also strongly prefer smaller semi-major axes within Earth's Hill sphere (\tab{tab:residenceTimePercentages}).  The prompt and delayed minimoons exhibit similar behavior in inclination and eccentricity and spend about 2/3 of their time on retrograde orbits and about 60\% of their time with geocentric eccentricity $e_g\ge0.5$ (\tab{tab:residenceTimePercentages}).

If the differences in the orbital distributions of TBOs with MB and lunar provenance is confirmed by future analyses it suggests that their orbital elements could provide an initial estimation of the likelihood that they derive from each source population.

The orbital element evolution of four known TBOs on  \fig{fig.residenceTimeDensity}\ allows a visual comparison to the TBOs with lunar provenance and two of them,\RH\ and \CD, both fulfill the minimoon conditions.  None of the objects reach the region of $(a,e,i)$ orbital element phase space with the highest residence time density, \ie\ $a\lesssim3$~LD, $e>0,5$ and $i>90\arcdeg$ which might be interpreted as an error in the simulations but it is more likely to be an observational selection effect.  The largest captured objects can be detected by the asteroid surveys when they are moving slowly, which favors the detection of large objects on distant orbits.  The far more common small objects on orbits with smaller semi-major axes and high eccentricity are difficult to detect.

\subsection{Earth impact rate after a lunar impact}
\label{ss.EarthImpactRateAfterLunarImpact}

In Neal Stephenson's book `Seveneves' the Moon catastrophically disrupts and lunar ejecta rain down on Earth (it also mentions minimoons).  Here we examine the impact rate on Earth after a less catastrophic asteroid impact on the Moon with special focus on a canonical impactor of $1\km$ diameter at their most likely impact speed of about $\OneKmLunarImpactSpeed\kps$ \citep{Marchi-2009}. 

\begin{center}
    \begin{minipage}{10cm}  
        \includegraphics[width=\columnwidth]{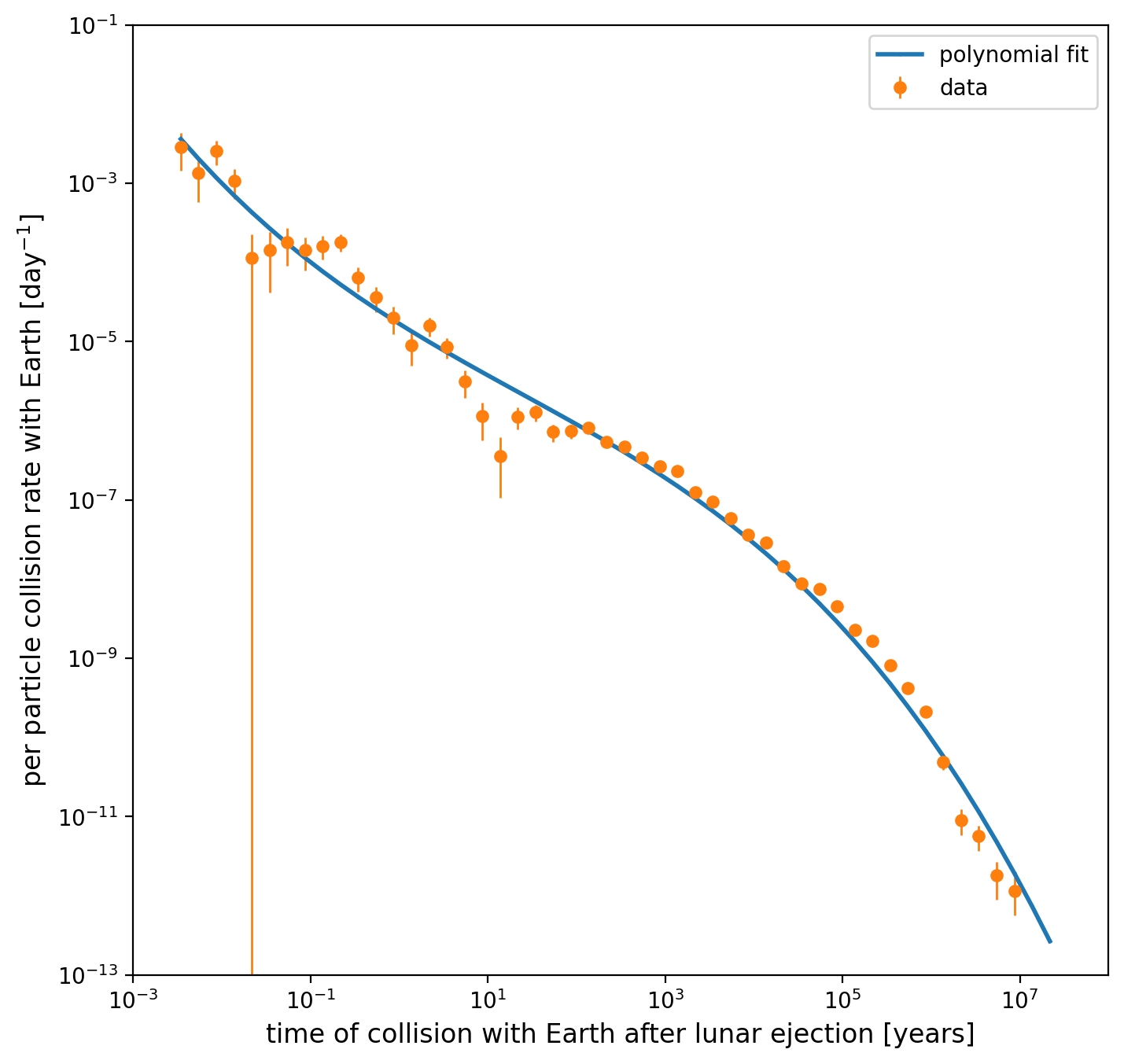}
        \captionof{figure}{
            The fraction of lunar ejecta that collide with Earth per day as a function of time since the ejection event. The fraction is zero until $\collisionMinTimeDays\Days$ after ejection.
        }
        \label{fig.fractionalEarthCollisionRate}
    \end{minipage}
\end{center}

We used the results of our lunar ejecta integrations (\S\ref{ss.LunarEjectaIntegrations}) to determine the fraction of ejected particles that collide with Earth per unit time, $f_{collide}(v,T)$, as a function of the ejection speed and time, $T$, after impact.  The fraction of ejected particles that collide with Earth in the time range $[T-\Delta T, T+\Delta T]$ is
\begin{equation}
    f_{collide}(v, T) \; \Delta T 
    = \frac{N_{collide}(v, T+\Delta T) - N_{collide}(v, T-\Delta T)}
           {N_{escape}(v)}.
\end{equation}
where $N_{collide}(v,T)$ is the cumulative number of particles launched at speed $v$ that collide with Earth for $t<T$ and $N_{escape}(v)$ is the total number of ejected particles at speed $v$.  

We found that $f_{collide}$ is consistent with being independent of the ejection speed, and the minimum time between lunar ejection and collision with Earth was $\collisionMinTimeDays\Days$ with a steep rise in collisions two days after ejection followed by a rapid drop-off.  Thus, we get
\begin{equation}
    f_{collide}(T) \; \Delta T = 
    \begin{cases}
        \frac{ \sum_k N_{collide}(v_k, T+\Delta T) - N_{collide}(v_k, T-\Delta T) }{ \sum_k n_{escape}(v_k) }, & \text{if } T \ge \collisionMinTimeDays\Days\\
        0,                 & \text{otherwise},
    \end{cases}
\end{equation}
where $v_k$ are the ejection speeds employed in our simulations.  There was also a pronounced $\sim27\Day$ periodicity in impacts during the first few months, presumably linked to the ejecta's lunar origin and the Moon's orbital period.

Collisions on Earth occur over the course of $10^7\yr$ (\fig{fig.fractionalEarthCollisionRate}), and the collision rate during that time decreases by about 10 orders of magnitude.  We fit the average fractional collision rate to a 3$^{rd}$ order polynomial with the result that
\begin{equation}
   y(T)  = (\fCollideParzero)  \; x(T)^3
         + (\fCollideParone)   \; x(T)^2
         + (\fCollidePartwo)   \; x(T)
         + (\fCollideParthree) \; 
    \label{eqn.f_collide}
\end{equation}
where $y(T) \equiv \log f_{collide}(T/\Days)$ and $x(T) \equiv \log (T/\yr)$.

Thus, the impact rate on Earth of ejecta with $d \ge d_{min}$ at a time $T$ after a lunar impactor of diameter $D$ striking the Moon at speed $V$ is simply
\begin{equation}
    r_{impact}(D, V, d_{min}, T) = f_{collide}(T) \; C(D,V) \; d_{min}^{-p_e}.
\end{equation}

\begin{center}
    \begin{minipage}{10cm}  
        \includegraphics[width=\columnwidth]{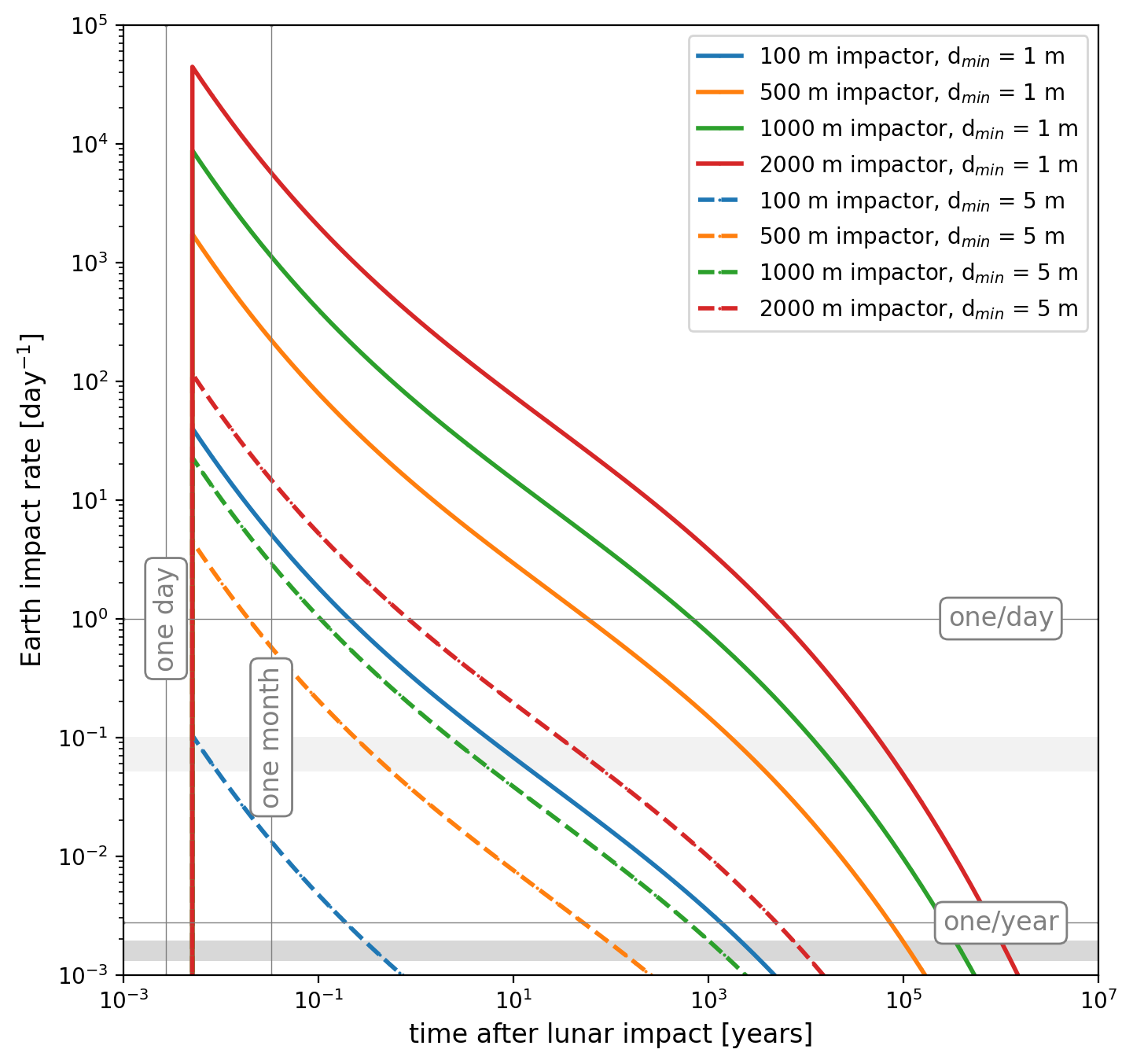}
        \captionof{figure}{
            The collision rate with Earth of $1\meter$ and $5\meter$ diameter lunar ejecta as a function of time since the ejection event and the lunar impactor's diameter at a constant speed of $\OneKmLunarImpactSpeed\kps$, the most likely speed for objects in the range $100\meter < D < 2\km$.  Thin gray horizontal lines indicate the once per day and once per year impact rates.  The light and dark shaded gray bands represent the measured range of impact rates on Earth of $1\meter$-diameter and $5\meter$-diameter bolides, respectively \citep{Brown2002,Brown2013}.
        }
        \label{fig.EarthCollisionRate}
    \end{minipage}
\end{center}

The rate of impact of $>1\meter$ diameter bolides on Earth after the impact of a $1\km$ diameter lunar impactor at $\OneKmLunarImpactSpeed\kps$ is about $1000\times$ the current rate \citep{Brown2002,Brown2013} for about $\daysThousandXCurrentOneMeterRate\Days$ after the lunar impact and remains above the current rate for about $\yearsGTCurrentOneMeterRate\yr$.  This suggests, with no surprise, that such an impact has not occurred within that amount of time.

\citet{Kreslavsky2014-DirectDeliveryofLunarImpactEjectatotheEarth} performed a similar study to ours but concentrated on the delivery of lunar material to Earth.  They found a maximum delivery rate of $>2$\%/day with a strong hemispherical and lunar latitudinal dependence.  Their maximum rate is about $7.5\times$ our maximum rate (\fig{fig.fractionalEarthCollisionRate}) but our value is averaged over the entire lunar surface and relies on different crater scaling functions and ejecta size-speed relations. They also provide the delivery rate per day, similar to \fig{fig.fractionalEarthCollisionRate}, for three specific lunar craters including Giordano Bruno that has a max delivery rate of about 1.5\%/day with a sharp distribution of Earth impacts $\sim3$ days after the lunar impact.  The Tycho and Thales craters produced max delivery rates of $\sim1$\%/day 2-3 days after impact with a longer tail of impacts lasting more than a week for the Thales crater.  The delivery rate falls exponentially in both their work and ours.

\citet{Jiao2024-Kamooalewa} suggest that asteroid \Kamo\ is a $36-60\meter$ diameter fragment of the Moon launched by the impact that created the $\sim22\km$ diameter Giordano Bruno crater $1-10\Myr$ ago.  Assuming a typical impact speed of $\OneKmLunarImpactSpeed\kps$ and employing the five crater scaling functions in this work (\tab{tab.CraterScalingSlopes}) with the associated nominal parameters, the Giordano Bruno impactor's diameter was in the range from $\GirodanoBrunoImpactorDiameterSinger\meter$ to $\GirodanoBrunoImpactorDiameterHH\meter$.  Thus, the current Earth impact rate of $1\meter$ scale objects from the impact of $2\km$ diameter object, in the upper range of sizes that would create the Giordano Bruno crater, is much less than the current rate of meter-scale impactors on Earth \citep{Brown2002,Brown2013}.  While the suggestion that Giordano Bruno was created in a lunar impact event only about 800 years ago \citep{Hartung1976-GiordanoBrunoObservedin1178} has been refuted \citep[\eg,][]{Withers.2001.EvidenceagainsttherecentformationoflunarcraterGiordanoBruno} this work suggests that, had it occurred, there would still be a few impacts per day of meter-scale objects on Earth, more than an order of magnitude higher than the current observed rate \citep[][and \fig{fig.EarthCollisionRate}]{Brown2013}.

\subsection{TBO capture rate after a lunar impact}
\label{ss.TBOCaptureRateAfterLunarImpact}

\begin{center}
    \begin{minipage}{10cm}  
        \includegraphics[width=\columnwidth]{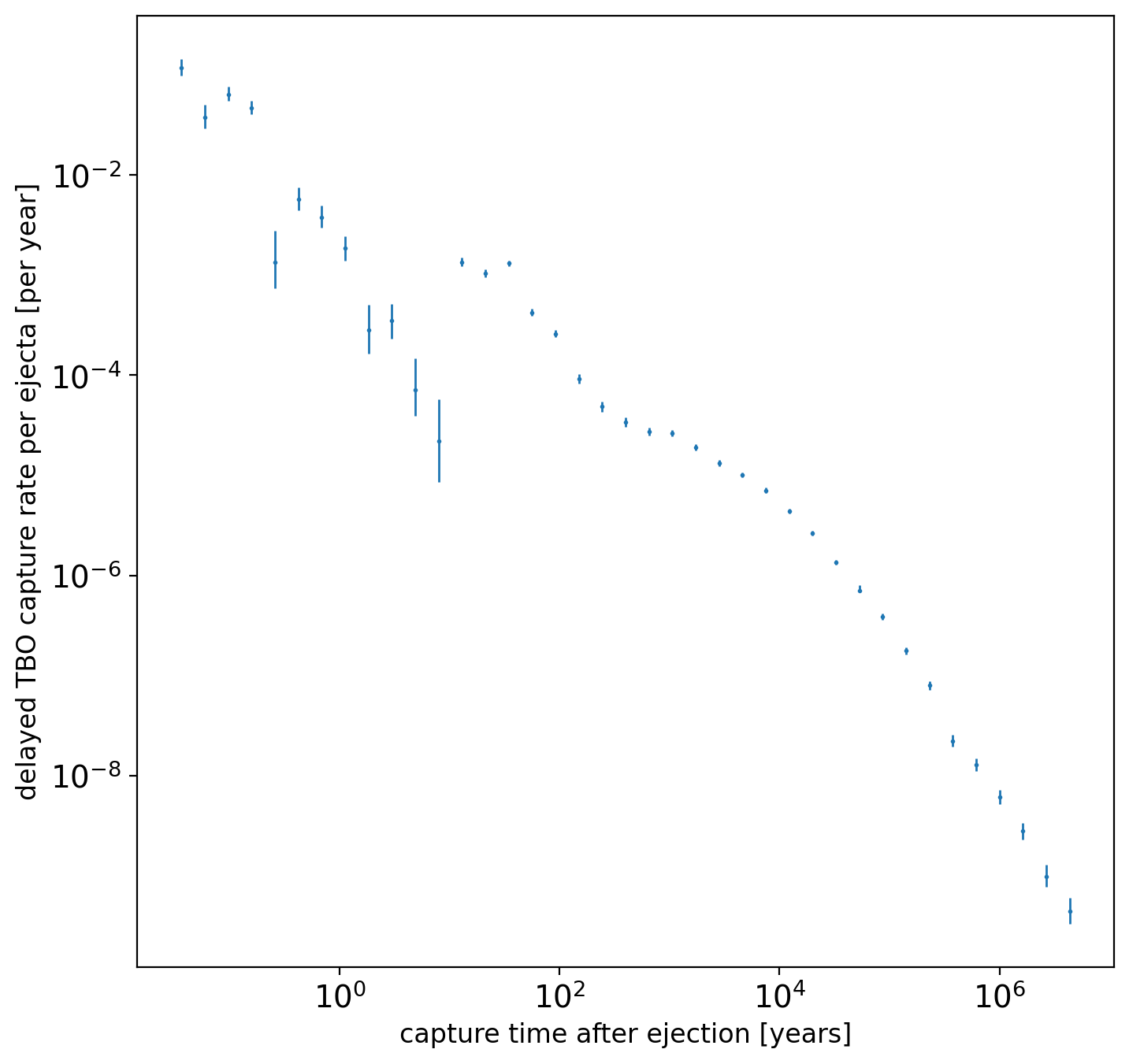}
        \captionof{figure}{
            The rate of delayed captures of lunar ejecta per ejecta as a function of time since the impact event.
        }
        \label{fig.TBOCaptureRateVsTime}
    \end{minipage}
\end{center}

After a lunar impact launches ejecta that becomes heliocentric and that may be captured in the EMS even millions of years afterwards, the capture rate of TBOs from \emph{that} event decays exponentially with time.  We used the results of our integrations (\S\ref{ss.LunarEjectaIntegrations}) to calculate the TBO capture rate as a function of time after the impact (\fig{fig.TBOCaptureRateVsTime}).  We were not able to calculate the rate as a function of ejection speed as in \S\ref{ss.EarthImpactRateAfterLunarImpact} because the number of delayed captures was too small, but we expect that the time spent in heliocentric orbit is sufficient to make the TBO capture rate insensitive to ejection speed, as was found for the Earth impact rate in \S\ref{ss.EarthImpactRateAfterLunarImpact}.

Once again, a canonical impactor of $1\km$ diameter at their most likely impact speed of about $\OneKmLunarImpactSpeed\kps$ \citep{Marchi-2009} ejects about $4\times10^6$ particles larger than $1\meter$ diameter (\fig{fig.ejectaSFD}).  Thus, \fig{fig.TBOCaptureRateVsTime}\ suggests that the TBO capture rate from such an impact decays to less than about one per year after about 10,000 years, but it must be remembered that the systematic uncertainty on this result is at least a few orders of magnitude (\S\ref{ss.minimoonSFD_systematics}).

\subsection{Heliocentric orbits for delayed TBOs}

\begin{figure}[htbp]
    \centering
    \includegraphics[width=0.48\columnwidth]{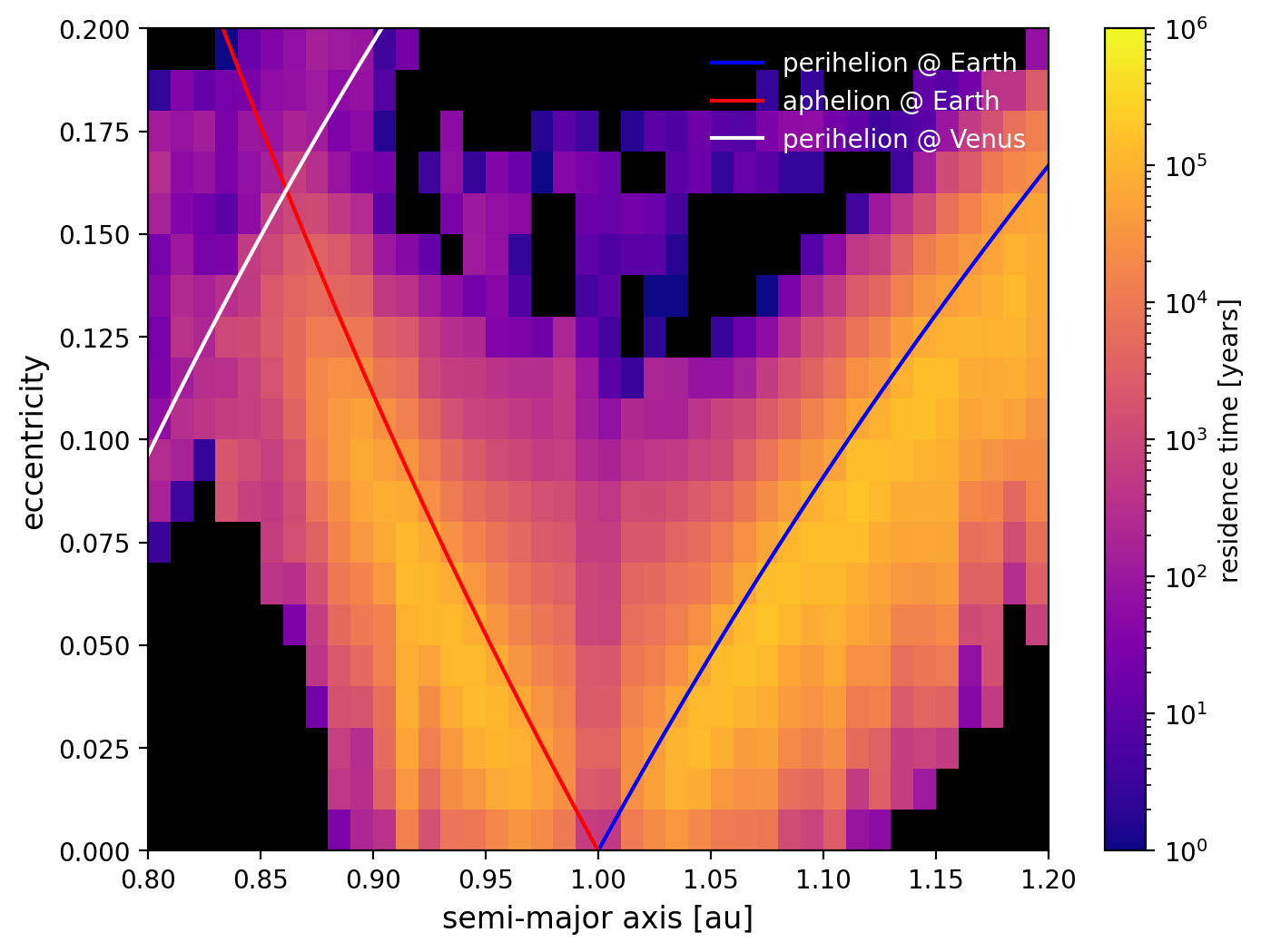}
    \includegraphics[width=0.48\columnwidth]{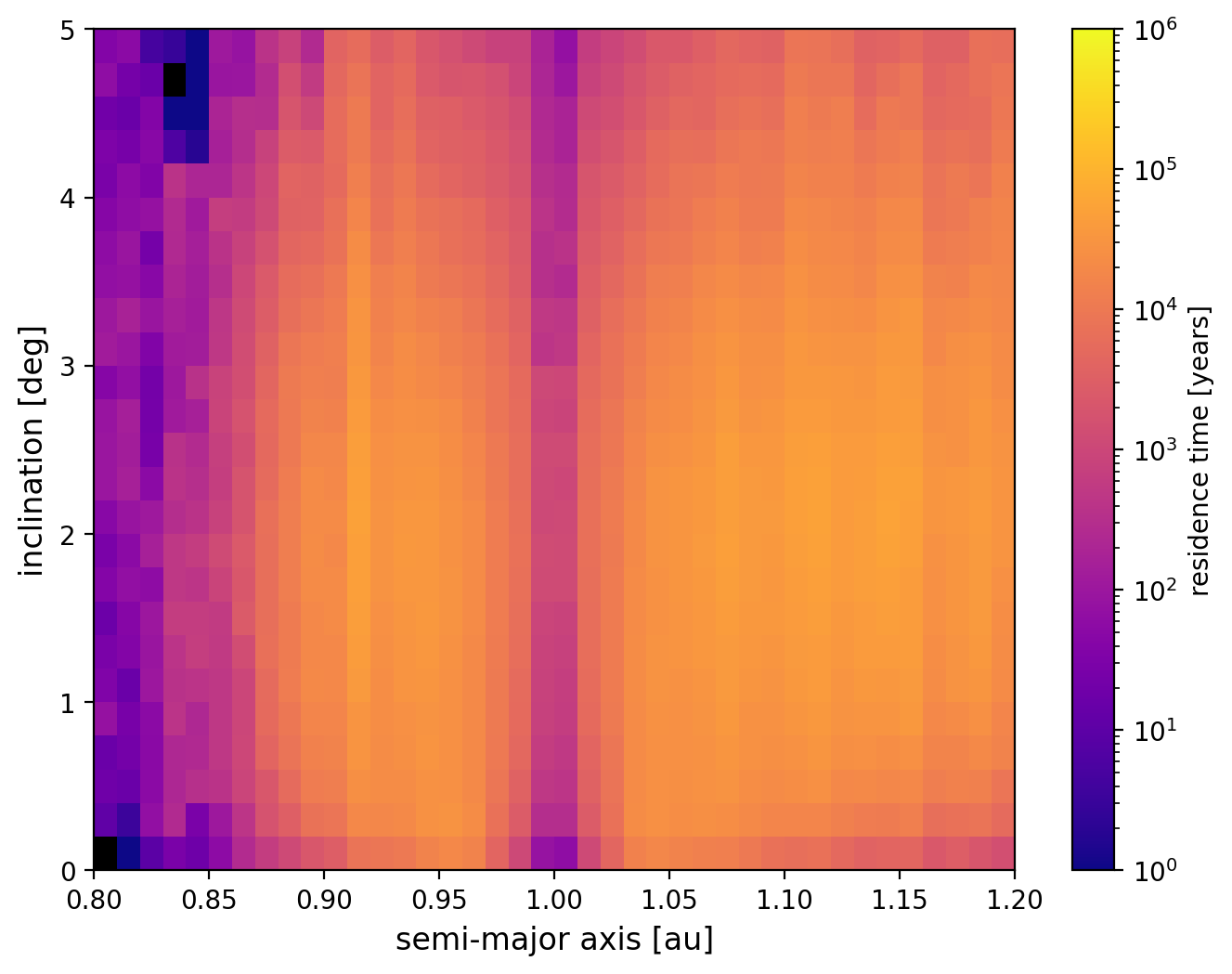}
    \caption{
        The heliocentric orbital element average residence time distribution per particle, the amount of time spent in each histogram bin, as a function of $(a,e)$ (left) and $(a,i)$ (right) for objects in the interval between their TBO periods.  The left panel includes curves corresponding to orbital elements with with perihelion or aphelion at $1\au$ or perihelion at Venus's semi-major of $0.723\au$. The bin widths are $\ResidenceTimeBinWidtha\au$, $\ResidenceTimeBinWidthe$ in eccentricity, and $\ResidenceTimeBinWidthi\arcdeg$.
    }
    \label{fig.heliocentricResidenceTime}
\end{figure}

We calculated a heliocentric residence time distribution, $p^{hel}_R$, for the particles when they were not TBOs (\fig{fig.heliocentricResidenceTime}) in a manner analogous to the geocentric residence time distribution (\S\ref{ss.GeocentricResidenceTimeDistributions}).

Our limits on the xISR, the heliocentric orbital element ranges from which we expected that delayed TBOs would occur, $0.8\au \le a \le 1.2\au$, $e\le 0.1$, and $i\le5\arcdeg$, were selected by more than doubling the ranges for minimoons employed by \citet{Granvik2012-minimoons} and \citet{Fedorets2017-minimoons} (see their fig. 8).  They found that temporarily-captured orbiters (TCO) and temporarily-captured flybys (TCF) were captured from heliocentric orbits in the ranges $0.93\au \lesssim a \lesssim 1.06\au$, $e\lesssim 0.075$, and $i\lesssim1.9\arcdeg$.  Thus, we were surprised to identify particles that were delayed TBOs outside their orbital element ranges (\fig{fig.heliocentricResidenceTime}).  Since the TBOs have much less stringent dynamical criteria than  minimoons they can become TBOs over a wider range of heliocentric $\aei$ than we allowed.

While there would be more residence time, $t_R$, if we had implemented a wider range of orbital elements in our xISR, \fig{fig.heliocentricResidenceTime}\ is visually misleading in its effect due to the logarithmic scale in the residence time.  We quantified the amount of `missed' residence time caused by truncating our heliocentric orbital ranges by linearly extrapolating the relationship between $t_R$ and both inclination and semi-major in the last few bins of both $a$ and $i$.  The missing residence time in semi-major axis amounts to $\sim\missingResidenceTimePercentSemiMajorAxis$\% of the total while the deficit due to the truncation in inclination is only $\sim\missingResidenceTimePercentInclination$\%.  The missing residence in the semi-major axis distribution is likely over-estimated by our linear extrapolation because the residence time appears to be falling exponentially as $a\rightarrow1.2\au$ so we expect that the actual neglected residence time amounts to $\sim10$\% of the total.

There is otherwise general agreement between the \citet{Fedorets2017-minimoons} residence time distributions for minimoons derived from the MB and our distributions for TBOs with a lunar origin.  Both works show a strong tendency for captures of objects that have perihelia or aphelia near $1\au$, roughly Earth's semi-major axis, a deficit of captured objects or residence time for $a\sim1\au$, and a reduced $t_R$ for objects with $e\sim0$ and $i\sim0\arcdeg$ due to the lack of actual objects in this phase space \citep{Harris2016-LifeAndDeathNearZero}.

The final minor difference between \citet{Fedorets2017-minimoons} and our work is that we observe a small `tail' in the $(a,e)$ residence-time distribution for objects dynamically evolving on orbits with perihelion at Venus (\fig{fig.heliocentricResidenceTime}), but this is not surprising given the well studied trajectories of ejecta traveling between the terrestrial planets \citep[\eg][]{Gault1983-TerrestrialAccretionofLunarMaterial,Gladman1996-ImpactEjectaBetweenTerrestrialPlanets}.

\subsection{Delayed capture TBO observability}

TBOs are difficult to detect due to their small sizes and rapid apparent rates of motion. \eg\ \citet{Fedorets2020-2020CD3} (fig. 5) suggests that \CD\ was only detectable by the Catalina Sky Survey (CSS) for about 2 nights of the $\sim1000$ nights that the $\sim2\meter$ diameter object was within the typical survey area after accounting for `trailing loss', the reduction in the signal per pixel due to the object's motion compared to a stationary object of the same apparent magnitude.  The CSS surveys a large fraction of the available night sky each clear night to a limiting stationary $V$ magnitude of $\sim21.5$ \citep{Christensen2019-CSS}, consistent with it successfully detecting \CD\ on one of the 2 nights that it was detectable.  Taken at face value, their results suggest a high detection efficiency if the trailed object is somewhere in the night sky and brighter than the system limiting magnitude, but it was only detectable on 0.2\% of the nights that it was a minimoon.

\begin{minipage}{15cm}  
    \includegraphics[width=\columnwidth]{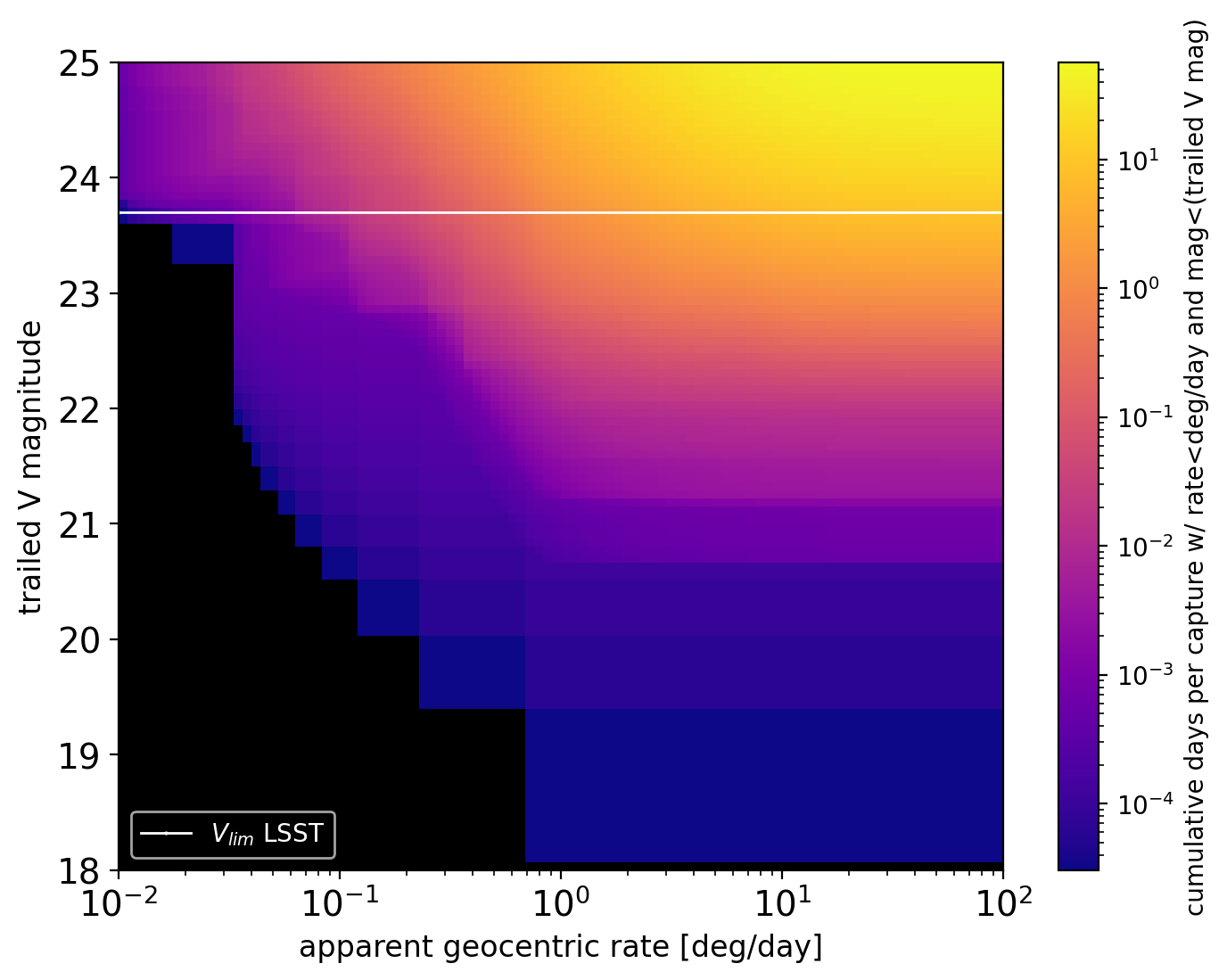}
    \captionof{figure}{
        The weighted cumulative average number of days of observability of $1\meter$ diameter TBOs of lunar provenance as a function of their rate of motion and trailed apparent magnitude in the VRO-LSST.  \ie\ at a given geocentric rate of motion and apparent magnitude the value on the figure is the average number density of days a lunar TBO with $H=32.75$, roughly $1\meter$ in diameter, is detectable in the VRO-LSST.  The horizontal line at trailed $V=\VlimLSST$ corresponds to the approximate limiting magnitude of the VRO-LSST in their regular survey mode \citep{Levine2024-ISC-LSSTselectionBiases}.
    }
    \label{fig.observability}
\end{minipage}

We calculated the apparent magnitude, $V$, and rate of motion, $\omega$, of all the delayed TBOs at each time step at each ejection speed, $v$, and determined the `observable residence time' of each object, \ie\ how long they are observable while $V_1 \le V_{trail} < V_2$ and $\omega_1 \le \omega < \omega_2$:
\begin{equation}
    t_{obs}(v; V_1 \le V_{trail} < V_2, \omega_1 \le \omega < \omega_2) = 
         \Delta t \frac{ \sum_{j=1}^{n(v)} \sum_{k=1}^{n_{times}} B(V_{jk};V_1,V_2) B(\omega_{jk};\omega_1, \omega_2) }{ n_{captures}(v) }
\end{equation}
where $B$ is the boxcar function defined in \S\ref{ss.GeocentricResidenceTimeDistributions}, we sum over all particles, $j$, ejected at speed $v$, and all time steps during which the particles were in a delayed TBO phase.  The total number of delayed TBOs at speed $v$, $n_{captures}(v)$, is different than the number of particles that experience a delayed TBO phase because particles may be bound more than once.

The apparent magnitude, $V$, was calculated using the full-precision reduced magnitude in the $H$-$G$ system \citep{Bowell1988-AsteroidsII} with $G=0.15$ and $H=32.75$ corresponding to a $1\meter$ diameter asteroid with an albedo of $0.143$.  The trailed magnitude, $V_{trail}$, at each apparent magnitude and rate of motion was calculated using the formulation described in \citet{Nesvorny2024-NEOMOD2}, which had also been employed in \citet{Fedorets2020-2020CD3}, with $V_{trail} = V + \Delta V$, where $V$ is the apparent magnitude, $\Delta V = 2.5 \log[ 1 + \LSSTTrailingInverseDPD ( \omega - \LSSTTrailingStartDPD ) ]$, $\omega$ is the apparent rate of motion in deg/day, and the other parameters are specific to VRO-LSST assuming it will deliver an average FWHM of $1''$ for the point-spread function with $\LSSTexpTime\sec$ exposures \citep[\eg][]{Schwamb2023-TuningLSSTforSSS}.  

Since the particles ejected at different speeds have different TBO capture fractions ($f(v)$, \S\ref{ss.captureFraction}), different probabilities of being delayed TBOs, we calculate the capture-fraction-weighted observable residence time:
\begin{equation}
    t^*_{obs}(V_1 \le V_{trail} < V_2, \omega_1 \le \omega < \omega_2) = 
        \frac{ \sum_v f(v) \; t_{obs}(v; V_1 \le V_{trail} < V_2, \omega_1 \le \omega < \omega_2) }
             { \sum_v f(v) }.
\end{equation}
Finally, we calculated the total average amount of time a delayed TBO would be observable with trailed apparent magnitude $<V_{trail}$ and rate of motion $<\omega$ (\fig{fig.observability}):
\begin{equation}
    T_{obs}(V,\omega) = \int^V \dif v \int^\omega \dif w \: t^*_{obs}(v,w).
\end{equation}

The VRO-LSST survey can detect one of the delayed TBOs for $\sim\TobservableLSST\Days$ when it has apparent rates of motion of $<\SurveyRateLimit$~deg/day.  The $\SurveyRateLimit\dpd$ rate limit is imposed by the LSST to reduce confusing natural with artificial objects that are roughly within about one lunar distance\footnote{personal communication with Grigori Fedorets, } (the Moon's angular rate is $\sim13\dpd$).  Since the LSST expects to image about half the sky at least every week it implies that the system will have about a 50\% detection efficiency for the lunar generated TBOs, but detection is not the same as discovery which recovers multiple detections of the same object.  The average delayed TBO duration of $\sim200\Days$ so the LSST should be able to detect a $1\meter$ diameter TBO for only a few percent of the time that it is bound.

\citet{Bolin2014} predicted that VRO-LSST could detect about 1.5 minimoons/lunation, almost 20/year, based on the population model of \citet{Granvik2012-minimoons} but assumed that the LSST would have a limiting $V$ magnitude of 24.7, a magnitude deeper than used here after accounting for the requirement that the system detect the object on 3 separate nights, and considered minimoons as small as $10\cm$ in diameter, corresponding to $H=37.75$.  They used a maximum detectable apparent rate of motion of $10\dpd$, like this work, but did not account for trailing losses.  \citet{Bolin2014}'s fig. 9 (left) suggests that there would be $\sim1$ minimoon larger than $1\meter$ diameter detectable by the LSST on the entire sky each night using the \citet{Granvik2012-minimoons} minimoon model.  This seems like an over-estimate given that the \citet{Granvik2012-minimoons} model claims that there is about $\sim1$ minimoon larger than $1\meter$ diameter at any time and those minimoons must spend some time in the direction of the Sun at phase angles that would make them undetectable to LSST.  

Our simulation suggests that the discovery of lunar-generated delayed capture TBOs is less likely than \citet{Fedorets2020-MinimoonsWithLSST} estimate that LSST might discover only 1 to 6 minimoons with a main belt provenance, objects that are dynamically similar to the the TBOs considered here. Given the small predicted discovery statistics by different groups it seems that LSST is unlikely to deliver a significant sample of TBOs and minimoons.

\subsection{Connections to the NEO population}

It has been more than 30 years since \citet{Rabinowitz1993-Arjunas} proposed that there is an `asteroid belt concentrated near Earth' which they termed `Arjunas' with an excess of objects of $<50\meter$ diameter `characterized by low eccentricities, widely ranging inclinations and unusual spectral properties'.  Their work prompted a series of papers over the ensuing decades studying whether the reported enhancement is real and attempting to ascertain the objects's provenance. \citet{Bottke1994-SEAProvenance} found that the objects were unlikely to be dynamically delivered from the MB but suggested that ejecta from the EMS or Venus, as studied in our work, could produce an excess of objects consistent with the Arjuna population. \citet{Michel+Froeschle.2000.smallEarthApproachers} agreed that the enhancement of small Earth approachers could not be explained by the dynamical evolution of objects from the MB but could be due to the collisional evolution and/or tidal splitting of Earth-crossing asteroids.  \citet{Brasser+Wiegert.2008-AsteroidsonEarth-likeorbits} then contradicted both of those earlier works by finding that asteroids on Earth-like orbits are more likely to be a sub-set of the NEO population that evolved from the MB than lunar ejecta. \citet{Granvik2024-NEOTidalDisruption} discovered that there is a detectable excess of objects on orbits tangential to the orbits of Earth and Venus, and showed that the orbital and size properties are consistent with tidal disruptions of NEOs during their close and slow encounters with Earth and Venus. In a series of papers including \citet{Nesvorny2024-NEOMOD2}, the authors debiased telescopic observations of asteroids by the CSS, independently confirmed that there is indeed an excess of small objects on Earth-like orbits that can not be explained with dynamical evolution of objects from the MB, and that tidal disruption is the only viable explanation.

Thus, the consensus is that there is an excess of asteroids on Earth-like orbits that is due to the tidal disruption of asteroids that have dynamically evolved from the MB onto Earth-like orbits.  Objects on Earth-like orbits are the minimoon/TBO source population and both \citet{Granvik2012-minimoons} and \citet{Fedorets2017-minimoons} showed that the population of NEOs on Earth-like orbits that evolved from the MB is consistent with minimoon statistics.  The NEO models upon which those two studies were based, \citet{Bottke2002a-NEOmodel} and \citet{Granvik2018-NEOmodel} respectively, did not explicitly include additional NEO sources such as tidal disruption, so their models most likely under-estimated the population of objects on Earth-like orbits.  This would imply that the minimoon/TBO population can be explained by the dynamical and physical evolution of MB asteroids.

This work finds that lunar ejecta due to asteroid and comet impacts \emph{could} also explain the minimoon/TBO population, but it is exquisitely sensitive to the uncertainties in crater formation (\S\ref{ss.minimoonSFD_systematics}).  Given that lunar ejecta could account for the TBO population, it could also contribute to the population of objects on Earth-like orbits with the same caveat.  This possibility is supported by recent evidence that some objects on Earth-like orbits have spectra more consistent with lunar basalt than any class of MB asteroid, \eg\ \Kamo\ \citep{Sharkey2021-2016HO3}, \CD\ \citep{Bolin2020-CD3}, \PT\ \citep{Bolin2024-PT5-arXiv}.  Future NEO population models that fit the debiased NEO population to linear combinations of various source populations, along the lines of those in the evolution from \citet{Bottke2002a-NEOmodel} to \citet{Granvik2018-NEOmodel} and then \citet{Nesvorny2024-NEOMOD3}, could incorporate lunar ejecta as a source population to predict the fraction of lunar material in the NEO population as a function of their diameter.

\section{Conclusions}

The steady-state population of Earth's TBOs, objects with a negative total energy with respect to the geocenter while within 3 Earth Hill radii, could be, at least in part, due to lunar material ejected from the Moon's surface after an asteroid or comet impact that subsequently evolve onto heliocentric orbit which are then re-captured by the EMS.  About 1 in 5 TBOs also satisfy the more stringent condition of being a minimoon. The problem is that predicting the size-frequency distribution of TBOs and minimoons derived from lunar impacts is uncertain by many orders of magnitude, due mostly to uncertainties in the crater formation process including the crater-scaling relations, the relationship between the impactor's impact energy and crater size, and the relationship between the ejecta's size and speed distributions.  The set of current and future objects on Earth-like orbits with Moon-like spectra could be used to determine the relative fraction of objects with a lunar or MB provenance and could constrain our understanding of the crater formation process.

\section*{Acknowledgements}

We thank Bryce T. Bolin, William F. Bottke, Gabriele Merli, Giovanni Valsecchi, and Tommaso del Viscio for their contributions and helpful discussions. 
We thank the two reviewers, Grigori Fedorets and the other anonymous, for their timely and thorough suggestions.
The integrations were performed on the Università di Pavia EOS HPC cluster. 
E.M. Alessi acknowledges  support from Fondazione Cariplo and the short-term mobility programme of CNR. M. Granvik acknowledges support from the Research Council of Finland (grant 361233).
Uncertainties on all the efficiencies and fractions in histograms and elsewhere were calculated using the technique and software developed by \citet{Paterno2004-EfficiencyUncertainty}.
This work made extensive use of the NumpPy \citep{NumPy}, SciPy \citep{SciPy}, and Astropy \citep{Astropy2013,Astropy2018,Astropy2022} packages.

\newpage

\bibliographystyle{abbrvnat}
\bibliography{references}

\newpage

\setcounter{figure}{0} \renewcommand{\thefigure}{A.\arabic{figure}}
\setcounter{table}{0} \renewcommand{\thetable}{A.\arabic{table}}
\setcounter{section}{0} \renewcommand{\thesection}{A.}
\setcounter{subsection}{0} \renewcommand{\thesubsection}{A.\arabic{subsection}}

\end{document}